\documentstyle[floats,preprint,aps,epsf]{revtex}
\tighten
\begin{document}
\draft


\title{Late time decay of scalar, electromagnetic, and gravitational
       perturbations outside rotating black holes}
\author{Leor Barack}
\address {Department of Physics,
          Technion---Israel Institute of Technology, Haifa, 32000, Israel}
\date{\today}
\maketitle


\begin{abstract}
We study analytically, via the Newman-Penrose formalism, the late time
decay of scalar, electromagnetic, and gravitational perturbations
outside a realistic rotating (Kerr) black hole.
We find a power-law decay at timelike infinity, as well as at null
infinity and along the event horizon (EH).
For generic initial data we derive the power-law indices for all radiating
modes of the various fields. We also give an exact analytic expression
(accurate to leading order in $1/t$) for the $r$ dependence of the late time
tail at any $r$.
Some of our main conclusions are the following.
(i) For generic initial data, the late time behavior of the fields is
dominated by the mode $l=|s|$ (with $s$ being the spin parameter),
which dies off at fixed $r$ as $t^{-2|s|-3}$ --- as in the Schwarzschild
background.
(ii) However, other modes admit decay rates slower than in the
Schwarzschild case.
(iii) For $s>0$ fields, non-axially symmetric modes dominate
the late time behavior along the EH. These modes oscillate along the
null generators of the EH.
\end{abstract}


\section{introduction}

Small perturbations of the Schwarzschild black hole (SBH) geometry die
off at late time with an inverse power-law tail.
This well-known phenomenon was discovered by Price early
in the 1970s.
Price explored the dynamics of linear scalar and metric perturbations
\cite{PriceI} (and that of all integer-spin fields in the Newman-Penrose
formalism \cite{PriceII}) propagating on the SBH background.
His analysis provided a detailed description of the relaxation mechanism
through which the black hole (BH) exterior settles down at late time
into its stationary ``no hair'' state.
In particular, Price was able to characterize the actual form of the
late time fall-off of the perturbations:
He found that any radiative multipole mode $l,m$ of an initially compact
linear perturbation dies off at late time as $t^{-2l-3}$ (where $t$ is
the Schwarzschild time coordinate). In the case there exists an initially
static multipole mode $l,m$ it will decay as $t^{-2l-2}$.
These power-law decay tails were found to be the same for all kinds of
perturbations, whether scalar, electromagnetic or gravitational (and
in this respect, the scalar field model proved to be a useful
toy model for more realistic fields).

Price's results were later reproduced using several different approaches,
both analytical and numerical
\cite{Leaver,GundlachI,Winicour,Andersson,BarackI,BarackII}, and were
generalized to other spherically symmetric BH spacetimes
\cite{GundlachI,Bicak,Ching,Burko,Brady,Contamination,Piran}. (A brief
review of the works on this subject can be found in the Introduction of
Ref.\ \cite{BarackI}.) The validity of the perturbative (linear) approach
was supported by numerical analyses of the fully nonlinear dynamics
\cite{GundlachII,Burko}, indicating virtually the same power-law indices
for the late time decay.

For a scalar field on the background of a SBH, power-law decay tails were
found to be exhibited also at future null infinity
\cite{Leaver,GundlachI,BarackII} and along the future event horizon
\cite{GundlachI,BarackII}. It was shown that at null infinity the scalar
field dies off with respect to retarded time $u$ as $u^{-l-2}$ (for a
compact initial mode) or as $u^{-l-1}$ (for a static initial mode). The
decay of the scalar perturbation along the event horizon (EH) was found to
be $v^{-2l-3}$ or $v^{-2l-2}$ (for a compact or a static initial mode,
respectively), where $v$ is the (Eddington-Finkelstein) advanced-time
coordinate.

As was already explained by Price, the late time tails of decay
outside spherically symmetric BHs originate from backscattering
of the outgoing radiation off spacetime curvature at very large distances.
In the framework of a frequency-domain perturbation analysis
\cite{Leaver,Andersson,Ching}, these tails are explained in
terms of a branch cut in the (frequency domain) Green's function,
which is, again, associated with the form of the curvature-induced
potential at large distance.
This suggests that the form of the decay at late time reflects (and
is affected by) merely the large distance structure of spacetime,
and may be independent of the existence or absence of an event horizon.
This assertion found further support in the analysis by Gundlach
{\em et al.\ }\cite{GundlachII}, who studied the
purely spherical collapse of a self-gravitating minimally coupled scalar
field. It was demonstrated numerically that in this case late
time tails form even when the collapse fails to create a black hole.
(On the other hand, quasi-normal ringing are found to dominate the early
stage of the waves' evolution only if a BH forms.)

Until quite recently, the issue of the late time decay of
BH perturbations has been considered only in spherically
symmetric models of BHs. It is known, however, that
astrophysically realistic BHs are spinning \cite{Bardeen},
and thus are not spherically symmetric but are rather of the
axially symmetric Kerr type. Moreover, it is suggested, in virtue of
the ``no hair'' principle \cite{Carter}, that the Kerr black hole
(KBH) might be the
{\em only} realistic BH (realistic BHs are not expected to
carry a significant amount of net electric charge).
Hence, generalization of the above mentioned analyses to the
KBH case seems to be of an obvious importance.
Still, such a generalization has awaited almost three decades,
as the lack of spherical symmetry in the Kerr background
makes both analytical and numerical exploration significantly more
complicated.

The ``no hair'' principle for BHs implies that perturbations
of the KBH must ``radiate away'' at late time. No
further information is available from this general principle
as to the actual details of this decay process.
A question arises as to what effect rotation has on the form of
the late time tails. More basically, does the decay of the
perturbing fields still obey a power law? If so, are the
power indices the same as in the SBH case?

Such questions were addressed only recently, by several authors.
First, Krivan {\em et al.} carried out a numerical simulation of
the evolution of linear scalar \cite{KrivanI} and gravitational
\cite{KrivanII} waves on the background of a Kerr black hole.
The first analytic treatment of this problem, for a scalar field,
was later presented by Barack and Ori \cite{Proc,Letter}
(following a preliminary analysis by Ori \cite{AmosI}).
Most recently, Hod used a different approach to study the late time
decay of both a scalar field \cite{HodI} and nonzero-spin
Newman-Penrose fields \cite{HodII,HodIII} on the Kerr background.
(Hod's analysis, which follows some preliminary considerations by
Andersson \cite{Andersson}, is carried out in the frequency domain,
whereas Barack and Ori use a time domain analysis.)
Finally, the analytic progress has motivated a further numerical
study by Krivan \cite{Rapid,Private}.

The above analyses all indicate that power-law tails of decay are
exhibited in the Kerr background as well.
In the Kerr case, however, the lack of spherical symmetry {\em couples}
between various multipole modes, which results in the
power-law indices of specific modes being found to be different,
in general, from the ones obtained in spherically symmetric BHs.
Another interesting phenomenon caused by rotation (first observed in
\cite{AmosI}) is the oscillatory nature of the late time tails
along the null generators of the EH of the Kerr BH for
nonaxially symmetric perturbation modes.

The purpose of the present paper is to extend the analysis described in
Ref.\ \cite{Letter} to electromagnetic and gravitational perturbations of
the Kerr background, and supply the full technical details of our
approach. (In Ref.\ \cite{Letter} we merely outlined the application of
our technical scheme to a scalar field, and gave a brief description of
the results in this case.) The analysis to be described in this paper
provides a more complete and accurate picture of the late time decay of
physical fields, than already available. Among the results which appear
here for the first time:\\ (i) We derive the form of the late time tail
for all radiative modes {\em anywhere} outside the KBH (i.e.\ at all
distances). We also give an exact analytic expression for the radial
dependence of this tail. (In \cite{HodI,HodII,HodIII} Hod only analyzes the
asymptotic behavior at very large distance, and along the EH.)\\ (ii) A
careful analysis of the decay along the EH reveals an interesting
phenomenon:  For $s>0$ fields, it is the oscillatory nonaxially symmetric
($m>0$) modes which dominate the late time behavior there. This result
has important implications to the
structure of the singularity at the inner horizon of the KBH \cite{Ori99}.
In a different paper \cite{Horizon} Barack and Ori further explore and
explain this phenomenon, and discuss the reason for the incorrect
prediction made in Ref.\ \cite{HodII} for the decay rate along the EH
in the $s>0$ case. (Recently \cite{HodIII}, following the appearance of
Ref.\ \cite{Horizon} as report no.\ gr-qc/9907085, Hod has corrected
his result.)

For simplicity, we shall refer in this paper only to the (most realistic)
situation in which the initial perturbation has a rather generic angular
distribution, such that it is composed of all multipole modes and, in
particular, the lowest radiatable one. It will be explained why an initial
setup in which the lowest modes are missing is more complicated
to explore using our approach. Hod's analysis \cite{HodI,HodII,HodIII}
provides predictions for this case as well; however,
these are not in agreement with recent numerical results by Krivan
\cite{Rapid}. It thus seems that further analytic work is needed
(using either Hod's method or ours) to clarify this point.

The arrangement of this paper is as follows.
In Sec.\ \ref{preliminaries} we briefly review the subject of
perturbations of
the Kerr geometry via the Newman-Penrose formalism. We then reduce
the master perturbation equation in the time domain to obtain a
coupled set of time-radial equations for the various modes. The
evolution problem for each of the modes is mathematically formulated as
a characteristic initial value problem.
In Sec.\ \ref{NullInfinity} we analyze the late time behavior of
the various modes at null infinity. To that end we apply
the {\em iterative scheme}, which is, basically, an extension
of a technique previously tested for a scalar field
in the SBH background \cite{BarackI,BarackII}.
In Sec.\ \ref{secLTE} we introduce the {\em late time expansion}
(LTE) scheme, which allows a global treatment of the decay at
late time. The late time behavior of fields at a fixed distance outside
the EH is dominated by the leading term of the late time
expansion, for which we derive an analytic expression in Sec.\
\ref{homogeneous}.
In Sec.\ \ref{EH} we then carefully explore the behavior along
the EH itself.
Finally, in Sec.\ \ref{Global}, we use the LTE scheme combined with
the results at null infinity, in order to derive the late time decay rates
for all modes of the fields at any fixed distance.
We conclude (in Sec.\ \ref{conclusions}) by summarizing our results
and discussing their relation to other works.

\section{Mode-coupled field equation} \label{preliminaries}

\subsection{Perturbations of the Kerr geometry via the Newman-Penrose
formalism (definitions and a brief review)}\label{subIIA}

The line element in Kerr spacetime reads, in Boyer-Lindquist (BL)
coordinates $t,r,\theta,\varphi$,
\begin{eqnarray} \label{Metric}
ds^2&=&-\left(1-2Mr/\Sigma\right)dt^2+\left(\Sigma/\Delta\right)dr^2
       +\Sigma d\theta^2
    +\left(r^2+a^2+2a^2Mr\sin^2\theta/\Sigma\right)\sin^2\theta\,
       d\varphi^2                                       \nonumber\\
    &-&\left(4aMr\sin^2\theta /\Sigma\right)d\varphi\, dt,
\end{eqnarray}
where $M$ and $a$ are, correspondingly, the BH's mass and specific
angular momentum, $\Sigma\equiv r^2+a^2\cos^2\theta$, and
\begin{eqnarray} \label{Delta}
\Delta\equiv r^2-2Mr+a^2.
\end{eqnarray}
(Throughout this paper we use relativistic units, with $c=G=1$.)
We shall consider in this paper only a BH solution with $|a|<M$:
the extremal case, $|a|=M$, requires a separate treatment, as we later
briefly explain.
The event and inner horizons of the (non-extremal) KBH are the two null
3-surfaces $r=r_+$ and $r=r_{-}$, respectively, where
\begin{equation} \label{rplus}
r_{\pm}=M\pm\sqrt{M^2-a^2}
\end{equation}
are the two roots of the ``horizons function'' $\Delta(r)$ defined in
Eq.\ (\ref{Delta}).

To discuss perturbations of the Kerr spacetime via the
Newman-Penrose (NP) formalism \cite{NP}, introduce Kinnersley's
null tetrad basis \cite{Kinnersley69}
$\left[l^\mu,n^\mu,m^\mu,m^{*\mu}\right]$ (where an asterisk denotes
complex conjugation).
In BL coordinates, the ``legs'' of this tetrad are given by
\begin{eqnarray} \label{Kinnersley}
l^\mu&=&\Delta^{-1}\left[r^2+a^2,\Delta,0,a\right] \nonumber\\
n^\mu&=&(2\Sigma)^{-1}\left[r^2+a^2,-\Delta,0,a\right] \nonumber\\
m^\mu&=&(2^{1/2}\bar{\rho})^{-1}
\left[ia\sin\theta,0,1,i/\sin\theta\right]
\end{eqnarray}
(with the fourth tetrad vector obtained from $m^{\mu}$ by
complex conjugation), where $\bar{\rho}\equiv r+ia\cos\theta$.
In the framework of the NP formalism, the
gravitational field in vacuum is completely described by five
complex scalars $\Psi_0,\ldots,\Psi_4$, constructed from the Weyl tensor
$C_{\alpha\beta\gamma\delta}$ by projecting it on the above tetrad basis.
Likewise, the electromagnetic field is completely characterized by the
three complex scalars $\varphi_0,\varphi_1,\varphi_2$, constructed by
similarly projecting the Maxwell tensor $F_{\mu\nu}$. In particular,
\begin{eqnarray} \label{Psi}
\Psi_0=-C_{\alpha\beta\gamma\delta}
           l^\alpha m^\beta l^\gamma m^\delta\quad\text{and}\quad
\Psi_4=-C_{\alpha\beta\gamma\delta}
           n^\alpha m^{*\beta} n^\gamma m^{*\delta}
\end{eqnarray}
represent, respectively, the ingoing and outgoing radiative parts
of the Weyl tensor, and
\begin{eqnarray} \label{varphi}
\varphi_0=F_{\mu\nu} l^\mu m^\nu\quad\text{and}\quad
\varphi_2=F_{\mu\nu} m^{*\mu}n^\nu
\end{eqnarray}
represent the ingoing and outgoing radiative parts
of the electromagnetic field.

In the (unperturbed) Kerr background all Weyl scalars
but $\Psi_2$ vanish (as directly implied by the Goldberg-Sachs
theorem in view of Kerr spacetime being of Petrov type D;
see Secs.\ 9b,9c in \cite{Chandra83}).
In the framework of a linear perturbation analysis,
the symbols $\Psi_0,\Psi_1,\delta\Psi_2,\Psi_3,\Psi_4$ and
$\varphi_0,\delta\varphi_1,\varphi_2$ are thus used to represent
first-order perturbations of the corresponding fields
(with $\delta\Psi_2\equiv \Psi_2-\Psi_2^{\rm{background}}$, etc.).
One can show (see Sec.\ 29b in \cite{Chandra83}) that
$\Psi_0$ and $\Psi_4$, and also $\varphi_0$ and $\varphi_2$,
are {\em invariant} under gauge transformations (namely, under
infinitesimal rotations of the null basis and infinitesimal
coordinate transformations).
The scalars $\Psi_1$ and $\Psi_3$ are not gauge invariant, and
may be nullified by a suitable rotation of the null frame.
The entities $\delta\Psi_2$ and $\delta\varphi_1$ represent perturbations
of the ``Coulomb-like'', non-radiative, part of the fields
(in fact, one can also nullify $\delta\Psi_2$ by a suitable infinitesimal
coordinate transformation.)
It is therefore only the scalars defined in Eqs.\ (\ref{Psi}) and
(\ref{varphi}) which carry significant information about the radiative
part of the fields.
(Note, however, that gauge invariance of the radiative fields
is guaranteed only within the framework of linear perturbation
theory.)

Teukolsky \cite{Teukolsky72} first obtained a single master perturbation
equation governing linear perturbations of scalar, electromagnetic,
and gravitational fields.
In vacuum, this master perturbation equation reads
\begin{eqnarray} \label{Master}
\lefteqn{\left[\frac{(r^{2}+a^{2})^2}{\Delta}-a^{2}\sin^{2}\theta\right]
\Psi^s_{,tt} - \Delta^{-s}\left(\Delta^{s+1} \Psi^s_{, r}\right)_{,r}
 +\frac{4Mar}{\Delta} \Psi^s_{,t\varphi}
+  \left(\frac{a^{2}}{\Delta}-\frac{1}{\sin^{2}\theta}\right)
\Psi^s_{,\varphi\varphi}}                          \nonumber\\
&&- \frac{1}{\sin\theta}\left(\Psi^s_{,\theta}\sin\theta\right)_{,\theta}
 -2s\left[\frac{a(r-M)}{\Delta}+\frac{i\cos\theta}{\sin^2\theta}
      \right]\Psi^s_{,\varphi}
 -2s\left[\frac{M(r^2-a^2)}{\Delta}-r-ia\cos\theta\right]\Psi^s_{,t}
                                                 \nonumber\\
&& +(s^2\cot^2\theta -s)\,\Psi^s  = 0.
\end{eqnarray}
Here, $\Psi^s(t,r,\theta,\varphi)$ represents the various radiative
fields according to the following list:
\begin{eqnarray} \label{fields}
\Phi   &=& \Psi^{s=0}\quad\text{(scalar field)},\nonumber\\
\varphi_0 &=& \Psi^{s=+1},                       \nonumber\\
\varphi_2 &=& (\bar{\rho}^*)^{-2}\Psi^{s=-1},    \nonumber\\
\Psi_{0}  &=& \Psi^{s=+2},                    \nonumber\\
\Psi_{4}  &=& (\bar{\rho}^*)^{-4}\Psi^{s=-2}.
\end{eqnarray}
The master equation (\ref{Master}) is fully separable only in the frequency
domain, by means of the {\em spin-weighted spheroidal harmonic} functions
$S^{slm}(-a^2\omega^2,\cos\theta)$ \cite{Teukolsky72}, where $\omega$ is the
temporal frequency (the separated equations are referred to as ``Teukolsky's
equation'').
Because the functions $S^{slm}(\theta)$ are $\omega$ dependent,
separation of the $\theta$ dependence is not possible in the
time domain (namely, without first decomposing the field into its
Fourier components).

\subsection{Reduction of the master field equation in the time domain}
\label{subIIB}

The target of the present work is to explore the behavior of the
fields $\Psi^s$ at late time.
In principle, the analysis can be carried out in the frequency
domain, as in Refs.\ \cite{HodI,HodII,HodIII}.
In this technique, the requested temporal behavior is finally to be
extracted by an inverse Fourier transform.
Our analysis is based on a different approach, motivated by the
following argument:
In the late time, stationary, limit ($t\rightarrow \infty$), one expects
the very low frequency ($\omega\rightarrow 0$) Fourier modes to dominate
the behavior.
For such waves, the functions $e^{im\varphi}S^{slm}(\theta)$ reduce to the
{\em spin-weighted spherical harmonic} functions $Y^{slm}(\theta,\varphi)$
\cite{Goldberg67}.
This may motivate one to try and extract the angular dependence of
the fields $\Psi^s$ by using the functions $Y^{lms}$.
As a result of the lack of spherical symmetry,
the resulting (time-domain) field equations will possess coupling between
the various multipole modes $l$; however, one should expect this
coupling to be ``small'', in a sense, at late time.
In the sequel we show how this coupling can be treated in an iterative
manner, in both the frameworks of the {\em iterative expansion}
(Sec.\ \ref{NullInfinity}) and the {\em late time expansion}
(Sec.\ \ref{secLTE}).

Led by the above argument, we expand the fields $\Psi^s$ as
\begin{eqnarray} \label{expansion}
\Psi^s(t,r,\theta,\varphi)=(r^2+a^2)^{-1/2}\Delta^{-s/2}
       \sum_{l=|s|}^{\infty}\sum_{m=-l}^{l}\,
       Y^{slm}(\theta,\varphi)\,\psi^{slm}(t,r),
\end{eqnarray}
where the radial factor in front of the summation symbols is
introduced for convenience [as it eliminates the term
$\propto\psi_{,r_*}$ from Eq.\ (\ref{basic1}) below].
Note that the summation over modes $l$ excludes
the $l<|s|$ modes, which are nonradiative (for a discussion regarding
the nonradiatable modes, see Sec.\ III\,D\,4 in Ref.\ \cite{PriceII}).

Inserting the expansion (\ref{expansion}) into the master field equation
(\ref{Master}), we obtain
\begin{equation} \label{NoSeparation}
\sum_{lm}\,Y^{slm}(\theta,\varphi)\left[\tilde{D}(t,r)\psi^{slm}
- a^2 \sin^2\theta\,(\psi^{slm})_{,tt}\right.
+ \left.2ias\cos\theta\,(\psi^{slm})_{,t}\right]=0,
\end{equation}
where $\tilde{D}(t,r)$ is a certain differential operator
independent of $\theta,\varphi$.
Note in Eq.\ (\ref{NoSeparation}) how the two last terms
in the squared brackets (proportional to $a$) avoid a full separation
of variables.

Now, the product $\cos{\theta}\cdot Y^{slm}$ can be re-expanded in
terms of the functions $Y^{slm}$ (which form a complete set of functions
on the unit 2-sphere for each $s$).
The ``matrix elements'' of the function $\cos\theta$ with respect
to the $Y^{slm}$ basis are given by \cite{Campbell71}
\begin{equation} \label{Clebsch}
\langle sl'm\vert \cos\theta\vert slm\rangle \equiv \oint
d\Omega\, (Y^{slm})^*\cos\theta\, (Y^{slm})
=\left(\frac{2l+1}{2l'+1}\right)^{1/2}\langle l1\,m0\vert l'm \rangle
\langle l1-s\,0\vert l'-s\rangle,
\end{equation}
where $\langle j_1j_2m_1m_2 \vert jm\rangle$ are the standard
Clebsch-Gordan coefficients \cite{AbramowitzStegun}.
We find that
\begin{equation} \label{cosYexpansion}
\cos\theta\cdot Y^{l}=
c_-^{l+1} Y^{l+1} + c_0^l Y^{l} + c_+^{l-1} Y^{l-1},
\end{equation}
where
\begin{equation} \label{cosYcoefficients}
\begin{array}{l}
c_-^l = \left[\frac{\left(l^2-s^2\right)\left(l^2-m^2\right)}
                       {l^2 (2l-1)(2l+1)}\right]^{1/2}, \\
c_0^l = -\frac{ms}{l(l+1)}, \\
c_+^l = c_-^{l+1}.
\end{array}
\end{equation}
(Here, as we shall often do below, we omit the indices $s,m$
for the sake of brevity.)
This also easily leads to
\begin{equation} \label{sin2Yexpansion}
-\sin^2\theta\cdot Y^l =
  C_{--}^{l+2} \, Y^{l+2}
+ C_-^{l+1}    \, Y^{l+1}
+ C_0^l        \, Y^{l}
+ C_+^{l-1}    \, Y^{l-1}
+ C_{++}^{l-2} \, Y^{l-2},
\end{equation}
where
\begin{equation} \label{sin2Ycoefficients}
\begin{array}{lcl}
    C_{++}^l &=&  c_+^{l+1} c_+^l,                      \\
    C_+^l    &=&  c_+^l (c_0^{l+1} + c_0^l),            \\
    C_0^l    &=& (c_-^l)^2+(c_+^l)^2+(c_0^l)^2-1,        \\
    C_-^l    &=&  c_-^l (c_0^l + c_0^{l-1}),            \\
    C_{--}^l &=&  c_-^{l-1} c_-^l .
\end{array}
\end{equation}
All constant coefficients $c^l$ and $C^l$
are nonvanishing, with only the following exceptions:\\
(i) $C_-^l$ and $c_-^l$ vanish for $l=|m|$ or $l=|s|$.\\
(ii) $C_{--}^l$ vanishes for $|m|\leq l\leq |m|+1$ or $|s|\leq l\leq
     |s|+1$.\\
(iii) $c_0^l$, $C_+^l$, and $C_-^l$ vanish for $m=0$ or $s=0$.\\
(It can be verified that $C_0^l$ is always negative definite.)

Substituting Eqs.\ (\ref{cosYexpansion}) and (\ref{sin2Yexpansion})
in Eq.\ (\ref{NoSeparation}) we obtain by the orthogonality of
the functions $Y^{slm}$,
\begin{equation} \label{psicoupled}
\bar{D}(t,r)\psi^l  +{\cal I}(\psi^{l\pm1},\psi^{l\pm2})=0
\end{equation}
for each $l,m,s$ satisfying $l\geq |m|$ and $l\geq |s|$,
where $\bar{D}$ is yet another differential operator,
and $\cal I$ is a functional describing {\em coupling}
between the $l$ mode and the $l\pm 1$ and $l\pm 2$ modes:
\begin{equation} \label{calI}
{\cal I}(\psi^{l\pm 1},\psi^{l\pm 2})=
a^2\left(C_{++}^l\psi^{l+2} +C_+^l\psi^{l+1}
+C_-^l\psi^{l-1}+C_{--}^l\psi^{l-2}\right)_{,tt}
+2ias\left(c_+^l\psi^{l+1} + c_-^l\psi^{l-1}\right)_{,t}.
\end{equation}
Note that, obviously, modes of different $m$ do not
interact with each other, as the Kerr geometry is axially symmetric.
We also point out that the scalar field case is special, in that
for this case the $l$ mode does not interact with the $l\pm 1$ modes,
but only with the $l\pm 2$ modes (recall that
the interaction coefficients $C_{\pm}^l$ vanish for $s=0$).

To write Eq.\ (\ref{psicoupled}) explicitly in a convenient
form, we introduce the advanced and retarded time coordinates,
defined, respectively, by
\begin{eqnarray} \label{EF}
v\equiv t+r_* \quad\text{and}\quad u\equiv t-r_*
\end{eqnarray}
(which are nonetheless {\em not} null coordinates in Kerr
spacetime).
Here, the ``tortoise'' radial coordinate $r_*$ is defined by
\begin{equation} \label{r*}
r_*= r-r_+ +(2\kappa_{+})^{-1}\ln z_+ -(2\kappa_{-})^{-1}\ln z_-,
\end{equation}
with $\kappa_{\pm}$ being the horizons' ``surface gravity'' parameters,
\begin{equation} \label{kappa}
\kappa_{\pm}=\frac{r_+-r_-}{4Mr_{\pm}},
\end{equation}
and where the dimensionless radial variables $z_{\pm}$ are given by
\begin{equation} \label{z}
z_{\pm}\equiv \frac{r-r_{\pm}}{r_+-r_-}.
\end{equation}
[Note the relation $\Delta=z_+z_-(r_+-r_-)^2$. Also, recall that we are
dealing in this paper only with non-extremal black holes, for which
$r_+>r_-$ and, consequently, $z_{\pm}$ are well defined.]
The coordinate $r_*$, satisfying $dr_*/dr=(a^2+r^2)/\Delta$, increases
monotonically with $r$ from $-\infty$ (at the EH) to $+\infty$ (at
spacelike infinity). Later we shall find useful the asymptotic relations
\begin{equation} \label{asy_r*}
\begin{array}{ll}
    e^{2\kappa_+ r_*}\simeq \frac{\Delta}{(r_+-r_-)^2}\simeq z_+ &
        \text {\ \  (for $r_*\ll -M$, $r\rightarrow r_+$)},\\
    r_*\simeq r &
        \text {\ \  (for $r_*\gg M$, $r\rightarrow \infty$)}.
\end{array}
\end{equation}

The explicit form of the mode-coupled field equation (\ref{psicoupled})
is now\footnote{
In Eq.\ (\ref{basic1}), the derivative $\partial_u$ is taken with
fixed $v$, $\partial_v$ is taken with fixed $u$, and $\partial_t$ with
fixed $r$.}
\begin{equation} \label{basic1}
\psi^l_{,uv}+ V^l(r)\psi^l + R^l(r)\psi^l_{,t}
+K(r)\left[a^2 C_0^l \,\psi^l_{,tt}
+{\cal I}(\psi^{l\pm1},\psi^{l\pm2})\right]=0,
\end{equation}
where $V^l(r)$, $R^l(r)$, and $K(r)$ are radial functions given by
\begin{mathletters}\label{potentials}
\begin{eqnarray} \label{V}
4 V^l(r)&=&(r^2+a^2)^{-2}\left[(l-s)(l+s+1)\Delta
-m^2a^2-2isma(r-M)\right]              \nonumber\\
&&-(r^2+a^2)^{-3/2}\Delta^{-s/2+1}\frac{d}{dr}
\left\{\Delta^{s+1}\frac{d}{dr}\left[(r^2+a^2)^{-1/2}
\Delta^{-s/2}\right]\right\},
\end{eqnarray}
\begin{equation} \label{R}
2R^l(r)=(r^2+a^2)^{-2}
\left[2imMar-sM(r^2-a^2)+s\Delta(r+iac_0^l)\right],
\end{equation}
\begin{equation} \label{K}
4K(r)=(r^2+a^2)^{-2}\Delta.
\end{equation}
\end{mathletters}

Of importance will be the large-$r$ asymptotic forms of these functions,
which in terms of the $1/r_*$ expansion read
\begin{mathletters}\label{AsyPotentials}
\begin{equation} \label{AsyV}
V^l(r)=\frac{l(l+1)}{4r_*^2}+
\frac{M-imsa+l(l+1)M\left[2\ln\left(\frac{r_*}{r_+-r_-}\right)-1\right]}
{2r_*^3} + O\left[\frac{(\ln r_*)^2}{r_*^{4}}\right],
\end{equation}
\begin{equation} \label{AsyR}
R^l(r)=\frac{s}{2r_*}+
\frac{s\left[iac_0^l-3M+2M\ln\left(\frac{r_*}{r_+-r_-}\right)\right]}
{2r_*^2} + O\left[\frac{(\ln r_*)^2}{r_*^{3}}\right],
\end{equation}
\begin{equation} \label{AsyK}
K(r)=\frac{1}{4r_*^2}-
\frac{M\left[1+2\ln\left(\frac{r_*}{r_+-r_-}\right)\right]}
{2r_*^3} + O\left[\frac{(\ln r_*)^2}{r_*^{4}}\right].
\end{equation}
\end{mathletters}

Equation (\ref{basic1}) already provides a qualitative picture of
one aspect of the fields' evolution:
Unlike in spherically symmetric spacetimes,
multipole modes of different $l$ interact with each other while
propagating on the Kerr background.\footnote{
See, however, our remark in the concluding section, with regard to the
definition of ``multipole moments'' in the Kerr geometry being somewhat
ambiguous.}
For example, if a physical ($s\neq 0$) perturbation is initially
composed of a pure $l,m$ mode, then, in general,
we may expect all possible modes $l',m$---namely, all modes with
$l'\geq \max(|m|,|s|)$---to be generated while the perturbation
evolves in time.
[The unrealistic case of a scalar field ($s=0$) is special, as for
this field only those of the above modes with even $l-l'$ are expected
to be excited.]
Note that all interaction terms in Eq.\ (\ref{basic1}) involve
derivatives of the field with respect to $t$.
It is this feature which, by means of the LTE scheme, allows one to
effectively decouple Eq.\ (\ref{basic1}), as we show in Sec.\
\ref{secLTE}.

\subsection{The initial setup}\label{subIIC}

We shall consider an initial perturbation in the form
of a compact outgoing pulse of radiation, which is relatively short,
yet arbitrarily shaped. We take this pulse to be emitted at
$u=u_0$, $v=v_0$, and without limiting the generality we take,
for simplicity, $v_0=0$. We further assume that the initial pulse
has a rather generic angular shape, so that it is composed of all
multipole modes $l,m$ [and in particular, for each $m$ it contains
the lowest possible mode, $l=\max(|m|,|s|)$].
Formulated mathematically, this initial setup takes the form\footnote{
It should be noted that, strictly speaking, these initial data
[supplemented to the coupled field equation (\ref{basic1})] do
{\em not} form a well posed characteristic initial-value problem,
as $u=$const and $v=$const are {\em not} characteristic hypersurfaces
of the Kerr geometry (these hypersurfaces are timelike rather than null).
We further comment on this issue below.}
\begin{equation} \label{initial}
\psi^{lm}=\left\{
\begin{array}{ll}
\Gamma^{lm}(u)& \quad\text{at $v=0$},\\
0             & \quad\text{at $u=u_0$},
\end{array}
\right.
\end{equation}
where, for each $l$ and $m$, $\Gamma^{lm}(u)$ is an arbitrary
(but nonvanishing) function with a compact support
between $u=u_0$ and (say) $u=u_1>u_0$, with $u_1-u_0\ll |u_0|$.
This type of initial data corresponds to the physical scenario in which
no ingoing radiation is coming from past null infinity.

It will be assumed in the following that the initial pulse is
emitted far away from the BH; namely, we take $-u_0\gg M$.
This assumption greatly simplifies our analysis (as we explain in
Sec.\ \ref{NullInfinity}; cf.\ \cite{BarackI,BarackII});
yet, it seems reasonable to expect the late time behavior in this case
to remain characteristic of the general situation.

\section{Late time behavior at null infinity: The iterative scheme}
\label{NullInfinity}

In this section we derive the form of the
late time decay at future null infinity, that is for $v\to\infty$ at
finite $u\gg M$. This has two main motivations:
First, the results at null infinity will appear to serve in the
framework of the late time expansion scheme as necessary
``boundary conditions'' for the global late time evolution
problem (as we discuss in Sec.\ \ref{secLTE}).
The results at null infinity also have their own physical significance,
for the following reason: Consider a static (fixed-$r$)
observer located at very large distance.
Let $\Delta u$ and $\Delta t$ represent, respectively, the retarded and
the static observer's time elapsed since this observer gets
the first signal from the perturbing field.
With respect to this observer, the relevant information
about the decay at finite $\Delta u=\Delta t$ is the one calculated
at the null infinity domain (which we may call, in this context, the
``astrophysical
zone'' of the waves). Only when the time lapse becomes infinitely large
(while $r$ remains finite) does this observer enters the ``future timelike
infinity'' zone, $t\gg r$ (the late time behavior in this domain will be
discussed in Sec.\ \ref{Global}).

For this part of the analysis (namely, for the derivation of the late
time tails at null infinity) we apply the {\em iterative scheme},
first developed and tested for scalar waves on the Schwarzschild
background in Refs.\ \cite{BarackI,BarackII}. Since the technical
details of the calculations involved are often very similar to
those in the above references, we mainly describe here the
new results (for spin-$s$ fields in Kerr), and direct
the reader to Refs.\ \cite{BarackI,BarackII} for further details.

\subsection{Formulation of the iterative scheme}\label{subIIIA}

We define
\begin{eqnarray} \label{V0R0}
V^l_0(r)&\equiv& \frac{l(l+1)}{4r_*^2},\nonumber\\
R_0(r)  &\equiv& \frac{s}{2r_*}
\end{eqnarray}
and
\begin{eqnarray} \label{DVDR}
\delta V^l(r)&\equiv& V^l(r)-V^l_0(r),\nonumber\\
\delta R^l(r)&\equiv& R^l(r)-R_0(r).
\end{eqnarray}
The functions $V_0^l(r)$ and $R_0(r)$ extract at large $r$ the (flat space)
asymptotic behavior of the functions $V^l(r)$ and $R^l(r)$ appearing in
Eq.\ (\ref{basic1}). The ``curvature induced'' residual part of these
functions is represented by $\delta V^l(r)$ and $\delta R^l(r)$.

We now decompose each of the functions $\psi^{slm}$ (for each $l,m$)
as
\begin{equation} \label{IE}
\psi^l=\psi^l_0+\psi^l_1+\psi^l_2+\cdots,
\end{equation}
such that each of the functions $\psi_n^l$ satisfies the field
equation
\begin{equation} \label{hierarchy}
\psi^l_{n,uv}+V^l_0(r)\psi^l_n+R_0(r)\psi^l_{n,t}=S^l_n,
\end{equation}
with $S^l_{n=0}\equiv 0$ and with
\begin{equation} \label{S}
S^l_{n>0}\equiv -\delta V^l\psi^l_{n-1}
                -\delta R^l(\psi^l_{n-1})_{,t}
-K(r)\left[a^2 C_0^l \,(\psi^l_{n-1})_{,tt}
+{\cal I}(\psi^{l\pm1}_{n-1},\psi^{l\pm2}_{n-1})\right].
\end{equation}
We take the initial conditions for the various functions
$\psi_n^l$ to be
$\psi_{n=0}^l=\psi^l$ and $\psi_{n>0}^l=0$ on
the initial surfaces $u=u_0$ and $v=0$.

Equation (\ref{hierarchy}), supplemented by the above initial data,
constitutes a hierarchy of characteristic initial data problems
for the various functions $\psi_n^l$. Formal summation over $n$
recovers the original evolution problem for $\psi^l$.
For each $n>0$ (and for all $l$), the function $\psi_n^l$ admits
an inhomogeneous field equation, with a source term depending only
on the functions $\psi_{n'<n}$ preceding it in the hierarchy, and with
the function $\psi_{n=0}^l$ satisfying a closed homogeneous equation.
This structure allows one, in principle, to solve for all functions
$\psi_n^l$ in an iterative manner: first for all modes $l$ of
$\psi_{n=0}^l$, which then serve as sources to $\psi_{n=1}^l$, etc.
In general, each function $\psi_n^l$ shall have sources
coming from the modes $l$, $l\pm 1$, and $l\pm 2$ of $\psi_{n-1}$.

Of course, the effectiveness of the proposed iteration scheme crucially
depends on its convergence properties.
In that respect, the scheme as formulated above may seem
problematic, because, while the ``zeroth order'' ($n=0$) field equation
well approximates the actual field equation (\ref{basic1}) at large
distance, it fails to do so at the highly curved small-$r$ region
(actually, the functions $V_0$ and $R_0$, as defined above, diverge
at $r_*=0$).

It is possible (by redefining $V_0$ and $R_0$ at small $r$) to
construct a more sophisticated iteration scheme that would account for
the small-$r$ region of spacetime as well --- as was
done in Ref.\ \cite{BarackII} for the case of a scalar field
in the Schwarzschild spacetime.
In that case, it was demonstrated \cite{BarackII}
that the small-$r$ details of the background geometry have merely a
negligible effect on the late time behavior at null infinity.
In the Schwarzschild case, we were able to greatly simplify the analysis
by considering a toy model of a thin spherically symmetric shell of
matter having a flat interior. In that model, the late time behavior
of the scalar field at null infinity turned out to well
approximate the actual behavior in a ``complete'' Schwarzschild
model (see \cite{BarackI,BarackII} for details).

The above results all indicate that the late time decay of the scalar
field is predominantly governed by the large-$r$ structure of the
Schwarzschild spacetime. (This conclusion stemmed already from
several previous works, as we mentioned in the Introduction.)
We shall assume in this paper that the same is also valid for
all fields $\Psi^s$ propagating on the Kerr background.
To be concrete, we will consider in this section, only for the sake
of the calculation at null infinity, a model in which the Kerr
interior geometry is replaced by a flat (Minkowski) manifold.
There is no way of smoothly attaching a flat interior to a Kerr
exterior through a thin spherically symmetric material shell
(as was done in \cite{BarackI} in the Schwarzschild case);
alternatively, we take this attachment to be made through a
material layer of a finite width. The external ``radius'' of this
layer should be of order of a few $M$.
Based on our experience with the Schwarzschild case, we expect the
details of our model at small distances (and, in particular, the internal
structure of the material layer) to merely have a negligible
affect on the behavior of the waves at null infinity, at $u\gg M$.

We emphasize that the above simplified model is adopted only for the
analysis at null infinity, where it greatly reduces the amount of
technical details one should deal with.
For the global analysis carried out in the rest of this paper,
the ``complete'' KBH geometry shall be considered.

\subsection{Zeroth order iteration term}\label{subIIIB}

By definition, the function $\psi_{n=0}^l$ obeys (for all $l$) the
homogeneous equation
\begin{equation} \label{HomEq}
\psi^l_{0,uv} + V_0^l(r)\psi_0^l + R_0(r)\psi^l_{0,t}=0,
\end{equation}
with the initial conditions $\psi_{n=0}^l(v=0)=\Gamma^l(u)$ and
$\psi_{n=0}^l(u=u_0)=0$. Equation (\ref{HomEq}) simply describes the free
propagation of the field $\psi_{n=0}^{sl}$ in Minkowski spacetime,
provided that $r_*$ is replaced with the radial Minkowski coordinate
$r$.

The general solution to this equation reads\footnote
{Note that the homogeneous equation (\ref{HomEq})
is invariant upon simultaneously transforming $u\rightleftharpoons v$
and $s\to -s$. The two sums in Eq.\ (\ref{HomGeneral}) constitute two
independent homogeneous solutions, which are obtained from each other
by this transformation.}
\begin{equation} \label{HomGeneral}
\psi_{n=0}^{\text{general}}=
\sum_{j=0}^{l-s}A_j^{sl}\frac{g_0^{(j)}(u)}{(v-u)^{l-j}}
+\sum_{j=0}^{l+s}A_j^{-sl}\frac{h_0^{(j)}(v)}{(u-v)^{l-j}},
\end{equation}
in which $g_0(u)$ and $h_0(v)$ are arbitrary functions (with their
parenthetical superindices indicating the number of differentiations),
and where $A^{ls}_j$ are constant coefficients given by
\begin{equation} \label{A}
A_j^{sl}=\frac{(2l-j)!}{j!(l-j-s)!}.
\end{equation}
The above initial data for $\psi_{n=0}^l$ uniquely determine a specific
solution for this function [but not for either of the functions $g_0(u)$
and $h_0(v)$ in separate).
For the outgoing pulse initial setup it is possible
to write this specific solution in a convenient form by taking
$h_0(v)\equiv 0$, in which case we have
\begin{equation} \label{psi0}
\psi_{n=0}^{l}=
\sum_{j=0}^{l-s}A_j^{sl}\frac{g_0^{(j)}(u)}{(v-u)^{l-j}},
\end{equation}
with the function $g_0(u)$ given by\footnote{
The calculation leading to Eq.\ (\ref{g0}) is a straightforward
generalization of the one described in detail in Ref.\ \cite{BarackI}
for $s=0$.}
\begin{mathletters}  \label{g0}
\begin{equation}
g_0^{l>s}(u)=\theta(-u)\times
\frac{l-s}{(l+s)!}\int_{u_0}^u (u/u')^{l+s+1}
(u-u')^{l-s-1}(-u')^s\Gamma(u')du',
\end{equation}
\begin{equation}
g_0^{l=s}(u)= \frac{(-u)^s}{(2s)!}\Gamma(u).
\end{equation}
\end{mathletters}
Here, $\theta(x)$ is the standard step function, taking the value $1$
or $0$ according to whether $x$ is positive or negative, respectively.
Equations (\ref{psi0}) and (\ref{g0}) describe how the function
$\psi_{n=0}^l$ can be constructed given any form $\Gamma(u)$ for
the compact initial outgoing pulse.

Eqution (\ref{g0}) implies that the wave $\psi_{n=0}^l$ is sharply ``cut off''
at retarded time $u=0$. This effect is due to ingoing and outgoint waves
destructively interfering with each other at the origin of coordinates
($r=0$) inside the flat region internal to the material layer, at retarded
times $u>0$. [For a scalar field in the ``complete'' Schwarzschild model we
found in Ref.\ \cite {BarackII} an exponential decay of the waves at $u>0$
rather than a sharp cutoff --- thus the main support of $\psi_{n=0}^l$
remains compact even in this more sophisticated
(and much more complicated) model.]

Obviously, since $\psi_{n=0}^l$ is strictly compact in retarded time
(it is supported only in the range $u_0<u<0$), it does not contribute
to the overall late time (large $u$) radiation at null infinity.
Rather, the function $\psi_{n=0}^l$ will serve [via Eq.\ (\ref{hierarchy})]
as a source to higher-order ($n\geq 1$) terms of the iteration scheme,
which will form the late time tail of decay, as we show below.

\subsection{Green's function of the iteration scheme}
\label{subIIIC}

Using the Green's function method, we formally have, for each
of the functions $\psi_{n>0}^l$,
\begin{equation} \label{psin}
\psi^l_n(v,u)=\int_{0}^{v}dv'\int_{u_0}^{u}du'
G^l(v,u;v',u')S^l_n(v',u'),
\end{equation}
which allows one, in principle, to calculate these functions one by one,
in an inductive manner.
Here, $G(v,u;v',u')$ is the time domain Green's function in Minkowski
spacetime, defined as satisfying the equation
\begin{equation} \label{eqG}
G^l_{,uv}+V^l_0(r)G^l+R_0(r)G^l_{,t}=\delta(u-u')\delta(v-v'),
\end{equation}
with the causality condition
\begin{equation} \label{causality}
G^l(u<u')=G^l(v<v')=0.
\end{equation}

To solve for $G$, we use a straightforward generalization of the
method used in Ref.\ \cite{BarackI}.
First one shows that for Eq.\ (\ref{eqG}) to be consistent with
the causality conditions (\ref{causality}), one must have
$G(v=v')=[(v'-u)/(v'-u')]^s$ and $G(u=u')=[(v'-u')/(v-u')]^s$.
This establishes a characteristic initial-value problem for
the Green's function at $u>u'$ and $v>v'$.
Then, with the help of Eq.\ (\ref{g0}), one can obtain
(see \cite{BarackI} for more details)
\begin{equation} \label{G}
G^l(v,u;v',u')=
\sum_{j=0}^{l-s}A_j^{sl}\,\frac{g^{(j)}(u;v',u')}{(v-u)^{l-j}}\,
 \theta(u-u')\, \theta(v-v')\, \theta(v'-u),
\end{equation}
where the function $g(u;v',u')$ is given by
\begin{equation} \label{g}
g(u;v',u')=\frac{1}{(l+s)!}\,
\frac{(v'-u)^{l+s}(u-u')^{l-s}}{(v'-u')^l},
\end{equation}
and where the $j$ differentiations of this function are with respect
to $u$.
The factor $\theta(v'-u)$ in Eq.\ (\ref{G}) is related, again, to
the presence of the origin of coordinates inside the flat internal
region.

Finally, a general comment should be made about the application
of the iterative scheme to the Kerr spacetime:
Here (unlike in the Schwarzschild case) the characteristic surfaces
of the iterative scheme (i.e.\ the surfaces of constant $u$ or
constant $v$, which are not null but rather timelike) do {\em not}
coincide with the actual characteristics of the Kerr geometry.
Thus, strictly speaking, the ``causality condition'' stated in Eq.\
(\ref{causality}) does not hold for the actual Green's
function in the Kerr spacetime.
However, the surfaces $u=$const and $v=$const do approach the actual
null characteristics of the Kerr background at large distances.
Recalling that the form of the late time radiation
at null infinity is shaped mainly during the propagation at large
distances, it is reasonable to expect that the above problem will not
crucially affect the validity of our results.

\subsection{First-order iteration term}\label{subIIID}

The first contribution to the late time (large-$u$) tail at null
infinity comes from the function $\psi_{n=1}^l$.
This contribution also turns out to be the most dominant one,
with those of the functions $\psi_{n>1}^l$ smaller by one or more
factors of $M/|u_0|$ (recall that in our model we have $|u_0|\gg M$).
It is therefore of special importance to analyze in detail the behavior
of $\psi_{n=1}^l$, and derive its late time form at null infinity,
as we shall do now.

The function $\psi_{n=1}^l$ is calculated from Eq.\ (\ref{psin}), with
$n=1$. Since we only look for the behavior at null infinity,
we take the limit $v\to \infty$ of this equation.
At that limit, the Green's function appearing in the integrand
is dominated by merely the $j=l-s$ term of the sum in
Eq.\ (\ref{G}):
\begin{equation} \label{Gscri+}
G^l(v\to \infty)\cong
\frac{(l+s)!}{(l-s)!}\, v^{-s} \left[g(u;u',v')\right]^{(l-s)}
\,\theta(u-u') \,\theta(v'-u)
\end{equation}
(where the derivatives of $g$ are with respect to $u$).

With Eq.\ (\ref{g}) we now obtain for $\psi_{n=1}^l$ at null infinity
(for each $l$),
\begin{equation} \label{psi1Integral}
\psi_{n=1}^l(u,v\to\infty)=\frac{v^{-s}}{(l-s)!}
\int_{u_0}^{0}du'\int_{u}^{\infty}dv'
\frac{\left[(v'-u)^{l+s}(u-u')^{l-s}\right]^{(l-s)}}{(v'-u')^l}
S_{n=1}^l(u',v')
\end{equation}
(where, again, the $l-s$ derivatives are with respect to $u$).
In this expression, the lower limit of the integration over $v'$
was set to $v'=u$ due to the factor $\theta(v'-u)$ appearing in the
Green's function, and the upper limit of the $u'$ integration
was set to $u'=0$ in view of the compactness of $\psi_{n=0}^l$
(implying the compactness of the source $S_{n=1}^l$ as well).

The source function $S_{n=1}^l$ is calculated from Eq.\ (\ref{S}),
with $n=1$.
It contains, in general, contributions from the modes
$l'=l,l\pm1,l\pm2$ of $\psi_{n=0}^{l'}$.
These contributions to $\psi_{n=1}^l$ [via Eq.\ (\ref{psi1Integral})]
are additive, and may be calculated one by one.
The details of this calculation are given in the Appendix.
In brief, it contains two steps for each of the above contributions:
First, the definite integration over $v'$ is  carried out explicitly.
The integrand of the remaining $u'$ integration then becomes a
finite sum over derivatives of $g_0^{l'}(u')$,
each multiplied by a power of $(u-u')$.
In the second and final step we use successive integrations by
parts to eliminate the derivatives of $g_0^{l'}(u')$, with all
resulting surface terms vanishing in virtue of the compactness of
this function. (This procedure is clarified in the Appendix.)

The following is a description of the outcome from the above calculation.
Let us denote by $\psi_{n=1}^{l'\to l}$ the contribution to the mode
$l$ of $\psi_{n=1}$ at null infinity from all terms in
$S_{n=1}^l$ associated with the mode $l'$ of $\psi_{n=0}$.
We then find at $u\gg -u_0$, to leading order in $M/u$ and in $u_0/u$
(see the Appendix for details),
\begin{equation} \label{contributions}
\begin{array}{lll}
\psi_{n=1}^{l-2\to l}&\cong&
        \alpha_{-2}^lI_0^{l-2}\,v^{-s}\,u^{-(l-s+2)},   \\
\psi_{n=1}^{l-1\to l}&\cong&
        \alpha_{-1}^lI_0^{l-1}\,v^{-s}\,u^{-(l-s+2)}\times \left[1+
        \beta_{-1}^l\ln\left(\frac{u}{r_+-r_-}\right)\right],\\
\psi_{n=1}^{l  \to l}&\cong&
        \alpha_{0}^l\,\,\,I_0^l\,\,\,\,\,\,v^{-s}\,u^{-(l-s+2)},\\
\psi_{n=1}^{l+1\to l}&\cong&
        \alpha_{+1}^lI_0^{l+1}\,v^{-s}\,u^{-(l-s+4)}\times \left[1+
        \beta_{+1}^l\ln\left(\frac{u}{r_+-r_-}\right)\right],\\
\psi_{n=1}^{l+2\to l}&\cong&
        \alpha_{+2}^lI_0^{l+2}\,v^{-s}\,u^{-(l-s+6)},
\end{array}
\end{equation}
where the $\alpha$'s and $\beta$'s are constant coefficients (depending
on $s,l,m$), and where $I_0^l$ is, for each $l$, a simple functional
of $g_0^l$:
\begin{equation} \label{I0}
I_0^l\equiv \int_{u_0}^0 g_0^l(u') du'.
\end{equation}
For a scalar field, $s=0$, the contributions $\psi_{n=1}^{l\pm 1\to l}$
vanish (as the coefficients $\alpha^l_{\pm 1}$ are proportional to $s$).
Also, there is no contribution $\psi_{n=1}^{l-1\to l}$ for $l<l_0+1$
and no contribution $\psi_{n=1}^{l-2\to l}$ for $l<l_0+2$,
where $l_0$ is the lowest radiatable multipole mode for given $s$ and
$m$:
\begin{equation} \label{l0}
l_0 = \max(|s|,|m|).
\end{equation}

Equation (\ref{contributions}) implies that for any given $s$ and $m$,
the most dominant mode of $\psi_{n=1}$ at null infinity, at large $u$,
is $l_0$.
It also tells us that the dominant contribution to this mode comes
solely from this mode itself (namely, it is $\psi_{n=1}^{l_0\to l_0}$),
since in this case there are no contributions from lower modes,
and the ones from higher modes are negligible at large $u$.
Hence, the decay of the dominant mode of $\psi_{n=1}$ at null infinity
(for any given $s$ and $m$) is described at large retarded time $u$
by
\begin{equation} \label{psi1l0}
\psi_{n=1}^{l_0}(u,v\to \infty)=
    \alpha_0^{l_0} \,I_0^{l_0}\,v^{-s}\,u^{-(l_0-s+2)},
\end{equation}
to leading order in $M/u$ and in $u_0/u$.
Recall that the function $g_0^{l_0}(u)$ and, hence, the functional
$I_0^{l_0}$ are nonvanishing,\footnote{
This is true unless the initial data are very finely tuned such as
to make the integral in Eq.\ (\ref{I0}) vanish.}
as long as we make the assumption that the initial
data have a generic angular form (so that it particularly contains the
lowest multipole mode $l_0$ for any given $s$ and $l$).
The case where the lowest modes are initially missing will be
discussed in brief below.

\subsection{Late time tail at null infinity}\label{subIIIE}

Equation (\ref{psin}) provides a formal means for calculating the higher
iteration terms, $\psi^l_{n\geq 2}$. However, exact analytic calculations
become very tedious already for the $n=2$ term. In the case of a scalar
field in Schwarzschild spacetime \cite{BarackI} we have explicitly derived
$\psi_{n=2}^l$ at null infinity, and showed that it exhibits the same
power-law decay as $\psi_{n=1}^l$ at large $u$, yet with an amplitude
reduced by a factor proportional to $(M/|u_0|)\ll 1$. For this case,
analytic considerations suggested that a similar reduction of the
amplitude by a factor $\propto M/u_0$ occurs also for $n>2$, whenever $n$
is increased by $1$. This conclusion was verified numerically for the
first few iterative terms \cite{BarackI}. Our numerical calculations also
indicated that the sum of iterative terms $\psi_n$ seems to converge
rather fast at null infinity for large $|u_0|/M$ (say, in the order of
$100$).

We now proceed under the assumption that the same considerations
also apply in our case, of general-$s$ fields on the Kerr
background. That is, we assume that (for any given $s,l,m$) the functions
$\psi_{n\geq 2}^l$ decay at null infinity at large $u$ with the same
tail as $\psi_{n=1}^l$; yet the amplitude of these functions is smaller
by at least one factor of order $O(M/u_0)$.
This assumption seems plausible, because the above property
of the iterative scheme [namely, the scaling of $\psi_n$ as
$\propto(M/u_0)^{n}$] seems to stem from the basic structure of the
iteration procedure, rather than from the details of the source function
$S_n$, which distinguishes the Schwarzschild, $s=0$, case from the
more complicated case studied in the current paper.

Adopting the above assumption, we conclude that, for large $|u_0|/M$,
the ``overall'' function $\psi^l$ is well approximated at null
infinity at large $u$ by merely the term $\psi_{n=1}^l$.
In particular, $l_0$ is the dominant mode of $\psi$ there, for
any given $s$ and $m$.
By virtue of Eqs.\ (\ref{expansion}) and (\ref{psi1l0}) we then finally
obtain for the Newman-Penrose field $\Psi^{sm}$ (for any given $s,m$),
\begin{equation} \label{Psil0}
\Psi^{sm}\cong
   \alpha_0^{l_0}\,I_0^{l_0}\,Y^{sl_0m}(\theta,\varphi)\, v^{-2s-1}
   \,u^{-(l_0-s+2)}\quad
   \text{(dominant mode at scri+, large $u$)}.
\end{equation}
This result is accurate to leading order in $M/u$, in $u_0/u$, and in
$M/u_0$.
In the generic case where the initial pulse includes all values of
$m$, we find that the behavior is dominated by the modes with
$0\leq |m|\leq |s|$ and $l=|s|$, which decay at null infinity with
the late time tail $u^{-2}$ for $s\geq 0$ or $u^{2s-2}$ for
$s\leq 0$.

One may also ask about the behavior of the other, faster decaying,
modes at null infinity.
From Eq.\ (\ref{contributions}) we find that (for any given
$s$ and $m$) the modes $l>l_0$ of $\psi_1$ are also ``fed'' by
strong contributions coming from modes of smaller $l$:
In general, the function $\psi_{n=1}^{l>l_0}$ has leading-order
contributions from the modes $l-1$ and $l-2$ of $\psi_{n=0}$.
In the exceptional scalar field case ($s=0$) this contribution
comes only from the $l-2$ mode (provided $l\geq l_0+2$), and
provides the same tail as the contribution from the
mode $l$ itseld, namely $u^{-(l-s+2)}$. We thus have, for all $l\geq l_0$
in the scalar field case, $\psi_{n=1}^l\propto u^{-(l-s+2)}$
at null infinity, large $u$.
Under the above assumption that the ``overall'' field $\psi^l$
is well approximated there by $\psi_1^l$ (for all $l$), we
conclude that {\em in the scalar field case}, the decay of {\em any}
of the modes $l\geq l_0$ at null infinity is given at large $u$ by
\begin{equation} \label{Psil}
\Psi^{lm}\propto
  Y^{lm}(\theta,\varphi)\, v^{-1}
   \,u^{-l-2}\quad \text{(any mode of a scalar field)},
\end{equation}
where $Y^{lm}$ are the spherical harmonics.

As to the non-dominating modes of the $s\neq 0$ fields: Equation
(\ref{contributions}) suggests that these modes would exhibit
not a strict power-law tail but rather a tail of the form
$\propto u^{-(l-s+2)}\times \ln(\frac{u}{r_+-r_-})$.
We feel, however, that this result cannot be taken as conclusive,
and needs a further support (e.g.\ from numerical analysis).
(We comment that such logarithmic dependence does not arise from
the frequency-domain analysis in Ref.\ \cite{HodII}.)
We emphasize our conclusion, Eq.\ (\ref{Psil0}), that the
leading-order tail at null infinity, belonging to the most
dominant mode, decays with a strict power-law.

\section{Late Time Expansion}\label{secLTE}

The target of this work is to explore the behavior of the
fields $\Psi^s$ at late time {\em anywhere} outside the KBH
(and along its EH).
To that end we shall apply the late time expansion scheme,
a version of which was used in Ref.\ \cite{BarackII} to analyze
a scalar field in the Schwarzschild case.

We assume that at late time, the fields $\Psi^s$ admit an
expansion of the form
\begin{equation} \label{LTE}
\Psi^s(v,r,\theta,\varphi)=\sum_{k=0}^{\infty}
\left[\sum_{l=|s|}^{\infty}\sum_{m=-l}^{l}\,
       Y^{slm}(\theta,\varphi)\,F_k^{slm}(r)\right] \, v^{-k_{0}-k},
\end{equation}
to which we shall refer as the {\em  Late Time Expansion} (LTE).
Here, $k_0$ is a constant parameter which we later
determine.\footnote{
It should be noted that {\em by definition} the parameter $k_0$ does
not depend on $l,m$: whereas in the Schwarzschild case \cite{BarackII}
a separate parameter $k_0^l$ has been defined for each mode $l,m$,
in the present paper a single parameter $k_0$ is related with the
overall field $\Psi^s$.}
As we show in this paper, the LTE is consistent with the field
equations, with the tail form at null infinity, and with regularity
requirements at the EH.
We adopt an expansion in $1/v$, rather than in $1/t$, because it
appears to be more adequate for analyzing the behavior near and
along the EH (as the coordinate $v$, unlike $t$, is regular through
the EH).

Inserting the form (\ref{LTE}) into the master field equation
(\ref{Master}), and collecting terms of common $v$ powers and of common
multipole numbers $l$ [with the aid of Eqs.\ (\ref{cosYexpansion}) and
(\ref{sin2Yexpansion})], yields an ordinary equation for each of the
unknown functions $F_{k}^{slm}(r)$:
\begin{equation} \label{BasicOrdinary}
D^{slm}\left[F_k^{slm}(r)\right]=Z_k^{slm},
\end{equation}
in which $D^{slm}$ is a differential operator given by
\begin{equation} \label{D}
D^{l}\equiv
\Delta \frac{d^2}{dr^2}+2(s+1)(r-M)\frac{d}{dr}
+\left[\frac{a^2 m^2 + 2isma(r-M)}{\Delta}-(l-s)(l+s+1)\right],
\end{equation}
and the source term $Z_k^{l}$ reads
\begin{eqnarray} \label{Z}
Z_k^{l} &=&
2(k_0+k-1)\left\{(r^2+a^2)\frac{dF_{k-1}^l}{dr}
+\left[\frac{2M[s(r^2-a^2)-imar]}{\Delta}+r-iasc_0^l\right]F_{k-1}^l
                                       \right.       \nonumber\\
&-&  ias\,\left(c_+^l F^{l+1}_{k-1} + c_-^l F^{l-1}_{k-1}\right)
                                                     \nonumber\\
&+& \left.\frac{a^2}{2}(k_0+k-2)\left(
C_0^l F_{k-2}^l + C_{++}^l F^{l+2}_{k-2} +C_+^l F^{l+1}_{k-2}
+C_-^l F^{l-1}_{k-2}+ C_{--}^l F^{l-2}_{k-2}\right)\right\},
\end{eqnarray}
with $F_{k<0}^l\equiv 0$, and with the various coefficients
$c^l$ and $C^l$ given in Eqs.\ (\ref{cosYcoefficients})
and (\ref{sin2Ycoefficients}).

An essential feature of Eq.\ (\ref{BasicOrdinary}) is the fact
that it actually constitutes a hierarchy of effectively {\em decoupled}
equations, as each of the functions $F_{k>0}$ satisfies an
inhomogeneous equation whose source depends only on the functions
$F_{k'<k}$ preceding it in the hierarchy. The first function,
$F_{k=0}^l$, obeys a closed homogeneous equation.
Thus, in principle, we may solve for all modes of all functions
$F_k$ one by one, starting with $k=0$. For each $k$, one should
be able to solve for all modes $l$ of $F_k$, and then carry on to $k+1$.

Now, Eq.\ (\ref{BasicOrdinary}) is a second-order
differential equation for each of the various functions $F_k^l$.
In principle, to determine these functions, proper boundary conditions
should be specified at the EH (which is, mathematically speaking,
a ``regular singular point'' of the equation) and at space-like infinity.
(These two boundary conditions should determine the two arbitrary
parameters which occur in the general solution for each of the
function $F_k^l$.)
The behavior of $\Psi^s$ at infinity is known from the previous section,
and in Sec.\ \ref{Global} below we discuss the matching to this
asymptotic region.
At the EH, the only obvious requirement concerns the regularity
of the physical fields there. In the rest of the present section
we obtain the boundary conditions for the functions $F_k^l$ at the
EH, based on local regularity considerations.

\subsection*{Boundary conditions at the EH} \label{subBC}

One expects ``measurable'' physical quantities to maintain a
perfectly regular behavior at the EH, which is a surface of
a perfectly regular local geometry.
Accordingly, the components of the Weyl and Maxwell tensors
should be perfectly smooth through the EH, provided these components
are expressed in a coordinate system regular at the EH.
To construct boundary conditions for the scalars $\Psi^s$ at the
EH, it then remains to relate these scalars to the regular components
of the Weyl and Maxwell tensors.

To that end we must first write the tetrad basis
(\ref{Kinnersley}), used to construct the scalars $\Psi^s$,
in EH-regular coordinates. Recalling that the BL coordinates
$t$ and $\varphi$ go irregular at the EH, we introduce the
Kruskal-like coordinates
\begin{equation} \label{Kruskal}
U\equiv -e^{-\kappa_+ u}\quad \text{and}\quad
V\equiv  e^{\kappa_+ v},
\end{equation}
and the regularized azimuthal coordinate
\begin{equation} \label{tildephi}
\tilde{\varphi}_+ \equiv \varphi - \Omega_+ t
\end{equation}
(see Sec.\ 58 in Ref.\ \cite{Chandra83}), where
\begin{equation} \label{Omega}
\Omega_+ = \frac{a}{2Mr_+}.
\end{equation}
($\Omega_+$ is the ``angular rate of inertial frame dragging'' at the EH.)
In the EH-regular coordinate system $(V,U,\theta,\tilde{\varphi}_+)$,
the components of the ingoing and outgoing tetrad legs have the
EH-asymptotic forms
\begin{equation} \label{tetradKruskal}
\begin{array}{lcl}
l^{\mu} & \propto & \Delta^{-1}\, e^{\kappa_+ v}
                            \left[1,0,0,0\right],\\
n^{\mu} & \propto & \Delta\, e^{-\kappa_+ v}\left[0,1,0,0\right].
\end{array}
\end{equation}
Recall that $v$ is regular at the EH, and that $\Delta=0$ there.

Now, the construction of the scalars $\Psi^{s}$ involves $|s|$
projections of the Weyl and Maxwell tensors on the
tetrad legs $l^{\mu}$ (for $s>0$) or $n^{\mu}$ (for $s<0$).
Since the components of these tensors must be EH-regular in the
coordinate system $(V,U,\theta,\tilde{\varphi}_+)$, then by virtue of
Eq. (\ref{tetradKruskal}) we find that
$\Delta^{s}\Psi^{s}$ must be EH-regular as well.

To formulate the regularity condition for the functions $F_k^l(r)$,
we note that the functions $Y^{slm}(\theta,\varphi)$ in Eq.\
(\ref{LTE}) are irregular at the EH for $m\neq 0$, due to the
factor $e^{im\varphi}$: We have
\begin{equation} \label{phi_irregular}
e^{im\varphi}=e^{im(\tilde{\varphi}_+ + \Omega_+ t)}=
\left[e^{im\tilde{\varphi}_+}\cdot e^{im\Omega_+ v}\right]
e^{-im\Omega_+ r_*},
\end{equation}
where the factor in square brackets is EH-regular, but the
following factor oscillates rapidly towards the EH.
From Eq.\ (\ref{LTE}) we thus find that it is the quantity
$\Delta^s e^{-im\Omega_+ r_*}F_k^l$
which must be perfectly regular at the EH (for all $k$).

Hence, the regularity condition at the EH can be phrased
as follows: Define the ``physical'' variables
\begin{equation} \label{Physical}
\hat\Psi^s   \equiv \Delta^s\Psi^s   \quad \text{and}\quad
\hat F_k^l   \equiv \Delta^s e^{-im\Omega_+ r_*}F_k^l.
\end{equation}
Then, we must have that (for all $k$)
\begin{equation} \label{BC}
\hat\Psi^s\,\,\,\text{and}\,\,\, \hat F_k^l\,\,\,
\text{are smooth functions at the EH.}
\end{equation}
Mathematically, we shall require these functions
and all their derivatives with respect to an EH-regular coordinate
(such as $r$ or $U$) to be continuous through the EH.

Equation (\ref{BC}) constitutes the required boundary condition at the
EH for all functions $F_k^l(r)$.

\section{Global solution for $\lowercase{k}=0$} \label{homogeneous}

The dominant late time decay at worldlines of fixed $r$ is
described by the $k=0$ term of the LTE, Eq.\ (\ref{LTE}).\footnote{
Actually, in Sec.\ \ref{EH} we discuss an exception to this statement:
For $m=0$ modes of $s>0$ fields, the behavior along the EH is
dominated by the $k=1$ term. However, in this case too,
the decay along lines of constant $r>r_+$ is still dominated
by the $k=0$ term.} In this section we derive an exact analytic
expression for this term (namely, for all modes $l$ of the function
$F_{k=0}^l)$. Since $F_{k=0}^l$ is a solution of a second-order
differential equation, it shall contain two arbitrary parameters.
One of these parameters will be determined in this section by the
regularity condition at the EH. The other parameter will be
determined in Sec.\ \ref{Global} through matching at infinity.

By definition, $Z_{k=0}^l=0$, and the function $F_{k=0}^l$ admits the
homogeneous equation
\begin{equation} \label{Static_Equation}
D^l\,F_{k=0}^l=0,
\end{equation}
with the operator $D^l$ given in Eq.\ (\ref{D}).
This is nothing but the {\em static field} equation in Kerr spacetime.
The static solutions play an important role in our analysis, for two
reasons: (i) As just mentioned, the late time behavior can be
approximated by knowing $F_{k=0}^l$, which must be a static solution, and
(ii) we shall use a basis of static solutions in constructing
the functions $F_{k>0}^l$ using the Wronskian method.

For reasons that will become clear below, we continue by treating
separately the cases $s\neq 0$ and $s=0$. Also, for $s\neq 0$ we will
consider separately the cases $m\neq 0$ (nonaxially symmetric modes)
and $m=0$ (axially symmetric modes).

\subsection{$s\neq 0$ fields: Nonaxially symmetric modes}
\label{subVA}

For $s\neq 0$ and $m\neq 0$, a basis of exact solutions to the homogeneous
static field equation (\ref{Static_Equation}) is given by
\begin{mathletters} \label{basis}
\begin{equation}
\phi_{r}=\left(z_+/z_-\right)^{im\gamma}
            \Delta^{-s}\, F(-l-s,l-s+1;1-s+2im\gamma;-z_+)
            \label{basis1},
\end{equation}
\begin{equation}
\phi_{ir}=\left(z_+/z_-\right)^{-im\gamma}
            F(-l+s,l+s+1;1+s-2im\gamma;-z_+), \label{basis2}
\end{equation}
\end{mathletters}
where $z_{\pm}$ are the dimensionless radial variables defined in
Eq.\ (\ref{z}),
\begin{equation} \label{gamma}
\gamma\equiv \frac{a}{r_+-r_-}=\frac{\Omega_+}{2\kappa_+},
\end{equation}
and $F$ denotes the {\em hypergeometric function}\cite{Chrzanowski75}.
(We use this notation because, as we discuss below, $\phi_r$ is
physically regular at the EH, whereas $\phi_{ir}$ is irregular
there.)

The hypergeometric function $F(\hat a,\hat b;\hat c;y)$ (where $\hat a$,
$\hat b$, and $\hat c$ are complex parameters and $y$ is a complex
independent variable) admits the series expansion
\begin{equation} \label{hypergeometric}
F(\hat a,\hat b;\hat c;y)=1+\sum_{n=1}^{\infty}
\frac{(\hat a)_n(\hat b)_n}{(\hat c)_n} \frac{y^n}{n!}
\end{equation}
(see, e.g., Sec.\ 2.1.1 in \cite{hg}), where
\begin{equation} \label{Pochhammer}
(\hat a)_n\equiv \hat a(\hat a+1)\cdots (\hat a+n-1)=
\Gamma(\hat a+n)/\Gamma(\hat a)
\end{equation}
is the ``rising factorial''.\footnote{
Note that $(\hat a)_n$ is well defined even when the expression
involving the gamma functions is not. In this case, of course, the
second equality in Eq.\ (\ref{Pochhammer}) is invalid.}
Two results arising from Eq.\ (\ref{hypergeometric}) are that
(i) the hypergeometric function is not defined if $\hat c$ is a
non-positive integer [as in this case a zero factor occurs in the
denominator in Eq.\ (\ref{hypergeometric})], and that (ii)
if either $\hat a$ or $\hat b$ are non-positive integers, the expansion
(\ref{hypergeometric}) terminates, and the hypergeometric function
becomes a polynomial of order $-\hat a$ or $-\hat b$, respectively.

Item (i) above implies that for $m\neq 0$ both solutions
(\ref{basis1}), (\ref{basis2}) are defined; however, in the case $m=0$
(which we treat separately below) only one of these solutions is defined
($\phi_r$ for $s<0$, or $\phi_{ir}$ for $s>0$).
We further find, by item (ii) above, that both hypergeometric functions
in the solutions (\ref{hypergeometric})  are
simply {\em polynomials} of $z_+$ (and thus of $r$ too).

For $m\neq 0$, the general static solution is constructed from the
two basis functions (\ref{basis}).
With the help of Eqs.\ (\ref{asy_r*}), (\ref{gamma}), and
(\ref{hypergeometric}), one finds the asymptotic forms of these
functions to be
\begin{equation} \label{basis1Asy}
\phi_r\cong \left\{
\begin{array}{ll}
    \Delta^{-s} e^{im\Omega_+ r_*} &
    \text{as $r_*\rightarrow -\infty$ ($\Delta,z_+\rightarrow 0$)}, \\
    B_s^{lm}\, r^{l-s}         & \text{as $r,z_+\rightarrow\infty$},
\end{array}
\right.
\end{equation}
and
\begin{equation} \label{basis2Asy}
\phi_{ir}\cong \left\{
\begin{array}{ll}
    e^{-im\Omega_+ r_*} &
    \text{as $r_*\rightarrow -\infty$ ($\Delta,z_+\rightarrow 0$)}, \\
    (B_{-s}^{lm})^{*}\, r^{l-s} & \text{as $r,z_+\rightarrow\infty$},
\end{array}
\right.
\end{equation}
where the coefficient $B_s^{lm}$ is given by
\begin{equation} \label{B}
B_s^{lm}=\frac{(2l)!\, \Gamma(1-s+2im\gamma)}
              {(l-s)!\, \Gamma(l+2im\gamma+1)}
              (r_+-r_-)^{-l-s}.
\end{equation}

We now use Eq.\ (\ref{Physical}) to construct the ``physical''
fields $\hat\phi_r\equiv  \Delta^s e^{-im\Omega_+ r_*}\phi_r$ and
$\hat\phi_{ir}\equiv  \Delta^s e^{-im\Omega_+ r_*}\phi_{ir}$
associated with the static solutions $\phi_r$ and $\phi_{ir}$.
From the EH-regularity criterion (\ref{BC}) we then learn that the
static solution $\phi_r$ is physically regular at the EH, whereas
$\phi_{ir}$ is irregular there: For $s<0$ the irregularity of
$\hat\phi_{ir}$ is obvious, as it diverges like $\sim \Delta^{-|s|}$
at the EH. For $s>0$, $\hat\phi_{ir}$ is continuous through
the EH, yet its $s$th derivative with respect to $U$ (which is a
regular coordinate through the EH) diverges there like
$\propto e^{-2im\Omega_+ r_*}$.
Higher order $U$ derivatives of $\hat\phi_{ir}$ are
unbounded in magnitude at the EH.

Obviously, $\phi_r$ is the only solution of the homogeneous static
equation (up to some global constant factor) which is physically regular
at the EH, because any combination $a\phi_r+b\phi_{ir}$ with
$b\neq 0$ will be irregular there.
In a physical setup where a static source presents outside the
BH, the field must behave as $\propto\phi_r$ near and through the EH.
In vacuum, $\phi_r$ is the only global static solution
physically regular at the EH. This field does not vanish at infinity
(where it behaves like $\propto r^{l-s}$);
hence it cannot represent a physical static perturbation.
There exists a static solution which dies off fast enough at
infinity [this is the solution
$(B_{-s}^{lm})^{*}\phi_r-B_s^{lm}\phi_{ir}$, which
dies off as $\propto r^{-l-s-1}$ at large $r$ \cite{Chrzanowski75}],
yet this solution is physically irregular at the EH.
Similar results apply also to the static $m=0$ modes, which we study
below.
We thus conclude that {\em there cannot exist physical vacuum static
modes} outside the Kerr BH. This, of course, is a manifestation of the
``no hair'' principle.

In the framework of the LTE, each of the functions $F_{k}^l(r)$ must
be subject to the EH-regularity criterion (\ref{BC}).
Since $F_{k=0}^l$ must be a static solution, it thus have to be
proportional to $\phi_r^l$.
To conclude the above discussion we therefore take
\begin{equation} \label{F0}
F_{k=0}^l(r) = a_0^l \,\phi_r^l(r).
\end{equation}
The constant $a_0^l$ is to to be determined in Sec.\ \ref{Global}
by matching at null infinity.

\subsection{$s\neq 0$ fields: Axially symmetric modes}
\label{subVB}

We remarked earlier that in the case $m=0$ only one of the basis functions
(\ref{basis1}), (\ref{basis2}) is defined---the one in which, depending on
the sign of $s$, the third parameter of the hypergeometric function is a
positive integer. Denoting this function by $\phi_r^{m=0}$, we have
\begin{mathletters} \label{phi0}
\begin{equation} \label{phi0a}
\phi_r^{m=0}=
F(-l+s,l+s+1;s+1;-z_+)\equiv \phi_r^+\quad  \text{for $s>0$}
\end{equation}
and
\begin{equation} \label{phi0b}
\phi_r^{m=0}=
\Delta^{-s}F(-l-s,l-s+1;-s+1;-z_+)\equiv \phi_r^-\quad
\text{for $s<0$}.
\end{equation}
\end{mathletters}
The asymptotic forms of this solution are
\begin{equation} \label{asyphi0EH}
\phi_r^{m=0}(r)\cong \left\{
\begin{array}{ll}
\Delta^0 & \text{(for $s>0$)} \\
\Delta^{-s} & \text{(for $s<0$)}
\end{array}\right.
\quad \text{at the EH}
\end{equation}
and, for both $s>0$ and $s<0$,
\begin{equation} \label{asyphi0INF}
\phi_r^{m=0}(r)\cong
B_{-|s|}^{l,m=0}\,r^{l-s}\quad
\text{as $r,z_+\rightarrow \infty$}.
\end{equation}
Recall that $F$ is simply a polynomial of $z_+$ (and of $r$), and thus so
is $\phi_r^{m=0}$.
From Eq.\ (\ref{phi0}) we find that the ``physical'' field
$\hat\phi_r^{m=0}\equiv \Delta^s \phi_r^{m=0}$
is also a polynomial, and therefore, clearly, $\phi_r^{m=0}$ is
physically regular at the EH.

We still have to construct a second independent basis static
solution for the $m=0$ case. (This will allow us to tell
whether or not the above solution $\phi_r^{m=0}$ is the only regular
one.)
Fortunately, at that point we can benefit
from the work already done in Ref.\ \cite{Horizon} for the Schwarzschild
case: When expressed in terms of the variable $z_+$ (rather than $r$),
the static field equation (\ref{Static_Equation}) for $m=0$ takes exactly
the same form as for the Schwarzschild BH [see Eq.\ (21) in \cite{Horizon}],
where in the latter case we use the variable $z\equiv (r-2M)/(2M)$.
Therefore, each static solution in the Schwarzschild spacetime
becomes a static axially symmetric ($m=0$) solution in Kerr
spactime, upon replacing $z\to z_+$.
Moreover, in terms of the variables $z$ (in the SBH case) and $z_+$
(in the KBH case), the EH-regularity criterion becomes the same for both
spacetimes, and thus the classification of regular and irregular solutions
at the EH is also conserved.

As a second basis function we then take the static solution
given in Eq.\ (24) of Ref.\ \cite{Horizon}:
\begin{mathletters} \label{eq13}
\begin{equation} \label{eq13a}
\phi_{ir}^{m=0}=
\tilde A_{ls}\,z_+^{-s}\,\,z_-^{-l-1}F(l-s+1,l+1;2l+2;z_-^{-1})
\equiv \phi_{ir}^+\quad                 \text{for $s>0$}
\end{equation}
and
\begin{equation} \label{eq13b}
\phi_{ir}^{m=0}=
\tilde A_{ls}\,z_-^{-l-s-1}F(l+s+1,l+1;2l+2;z_-^{-1})
\equiv \phi_{ir}^-\quad                 \text{for $s<0$},
\end{equation}
\end{mathletters}
where we have replaced $z$ with $z_+$ (and thus $z+1$ with $z_-$).
Here, $\tilde A_{sl}$ is a normalization factor,
\begin{equation} \label{tildeA}
\tilde A_{sl}=1/F(l-|s|+1,l+1;2l+2;1)
             =\frac{l!(l+|s|)!}{(2l+1)!(|s|-1)!}
\end{equation}
[cf.\ Eq.\ (46) in Sec.\ 2.8 of \cite{hg}],
chosen such that $\phi_{ir}^{m=0}$ takes a simple asymptotic form at
the EH (see below).
From Ref.\ \cite{Horizon} we also know that $\phi_{ir}^{m=0}$ admits
the following series expansion near the EH:
\begin{equation} \label{eq45}
\phi_{ir}^{m=0}(r)= \left\{
\begin{array}{ll}
\Delta^{-s}(1+\tilde\alpha^+ \Delta+\cdots)
+ \tilde\beta\,\phi_r^+\ln z_+      & \text{(for $s>0$)}, \\
(1+\tilde\alpha^- \Delta+\cdots)
+ \tilde\beta\,\phi_r^-\ln z_+      & \text{(for $s<0$)},
\end{array}\right.
\end{equation}
in which the coefficient $\tilde\beta$ is nonvanishing,
\begin{equation} \label{beta}
\tilde\beta=\frac{(-1)^{s+1}(l+|s|)!}{(|s|-1)!(|s|)!(l-|s|)!}
\,(r_+-r_-)^{-2|s|}.
\end{equation}

It is clear (e.g.\ by comparing the EH-asymptotic forms) that the
two solutions $\phi_r^{m=0}$ and $\phi_{ir}^{m=0}$ are independent,
and thus form a complete basis of solutions.
As already explained in Ref.\ \cite{Horizon}, $\phi_{ir}^{m=0}$ is
physically irregular at the EH: For $s<0$ the ``physical'' field
$\hat\phi_{ir}^{m=0}$ diverges there as $\Delta^{-|s|}$. For $s>0$
$\hat\phi_{ir}^{m=0}$ is continuous through the EH, yet its $s$th
derivative with respect to $U$ diverges there (as $\propto\ln z_+$).
Therefore, Eq.\ (\ref{F0}) applies to $m=0$ as well, where
for this case the function $\phi_r$ is given by Eq.\ (\ref{phi0}).

It is instructive to compare between the asymptotic behavior of the
$s\neq 0$ static solutions in the case $m\neq 0$
[Eqs.\ (\ref{basis1Asy}) and (\ref{basis2Asy})]
and in the case $m=0$ [Eqs.\ (\ref{asyphi0EH}) and (\ref{eq45})].
Focusing on the asymptotic dependence on $\Delta$ (and ignoring
for this discussion the oscillatory factor presents in the
$m\neq 0$ case),
we find that for the $s>0$ fields the regular and irregular
solutions ``switch roles'': For $m\neq 0$ modes the regular
solution is the one that behaves like $\Delta^{-s}$ at the EH and
the irregular solution is the one that behaves like
$\propto{\rm const}$
there, whereas for $m=0$ modes the opposite is true.
Such an interchange of roles does not occur in the case $s<0$.
This effect is explored and explained in detail in Ref.\
\cite{Horizon}

\subsection{Scalar field case ($s=0$)}
\label{subVC}

For $s=0$ we use the new radial variable
\begin{equation} \label{barz}
\bar z\equiv z_++z_-= \frac{2r-r_+-r_-}{r_+-r_-}
\end{equation}
(note the relation $\bar z=2z_++1=2z_--1$),
to write the static field equation (\ref{Static_Equation})
in the form
\begin{equation} \label{StaticScalar}
(1-\bar z^2)F''(\bar z)-2\bar z F'(\bar z)+
\left[4m^2\gamma^2(1-\bar z^2)^{-1}+l(l+1)\right] F(\bar z) = 0,
\end{equation}
where a prime denotes $d/d\bar z$.
This is the familiar Legendre's differential equation
(see, for example, Sec.\ 3.2 in Ref.\ \cite{hg}).
Two independent solutions to this equation are \cite{hg}
\begin{mathletters}\label{phiscalar}
\begin{equation}\label{P}
\phi_r^{s=0}=(z_+/z_-)^{im\gamma}\,F(-l,l+1;1+2im\gamma;-z_+)
\end{equation}
and
\begin{equation}\label{Q}
\phi_{ir}^{s=0}=\frac{(\bar z^2-1)^{im\gamma}}{\bar z^{l+2im\gamma+1}}
\,F(l/2+im\gamma +1,1/2+im\gamma +1/2;l+3/2;\bar z^{-2}),
\end{equation}
\end{mathletters}
which are (up to a customary normalization) the {\em associated
Legendre functions} of the first and second kinds,
$P_l^{\mu}(\bar z)$ and $Q_l^{\mu}(\bar z)$,
respectively, with $\mu=2im\gamma$.\footnote{
Note that the two independent solutions $\phi_r^{s\neq 0}$ and
$\phi_{ir}^{s\neq 0}$, Eqs.\ (\ref{basis1}) and (\ref{basis2}),
degenerate to the single solution (\ref{P}) in the scalar field
case (i.e., when setting $s=0$).}

Using the EH-asymptotic relations
$z_+\simeq e^{2\kappa_+ r_*}=e^{\Omega_+r_*/\gamma}$
and $z_-\simeq 1$, and
recalling that the hypergeometric function appearing in
Eq.\ (\ref{P}) is a polynomial (of order $l$) of $z_+$,
we find $\phi_r^{s=0}\simeq e^{im\Omega_+r_*}$ near the EH.
The ``physical'' field
$\hat \phi_r^{s=0}\equiv e^{-im\Omega_+r_*}\phi_r^{s=0}$
associated with this solution is therefore regular at the EH.

It remains to verify that $\phi_{r}^{s=0}$ represents the {\em only}
physically regular solution for the scalar field (up to a
constant factor).
For $m\neq 0$, the solution $\phi_r^{s=0}$ and its complex
conjugate $(\phi_r^{s=0})^*$ constitute a complete basis of
static solutions. The ``physical'' field
$e^{-im\Omega_+r_*}(\phi_r^{s=0})^*$
becomes indefinite at the EH (where it behaves like
$e^{-2im\Omega_+r_*}$), and we find that for $m\neq 0$,
$\phi_r^{s=0}$ is indeed the only physically regular solution.
In the case $m=0$, $\phi_r^{s=0}$ becomes real (it is then
the Legendre polynomial, up to a normalization), and a complete
basis of solutions is given by $(\phi_r^{s=0},\phi_{ir}^{s=0})$.
These two basis functions then admit a relation of the form
$\phi_{ir}^{s=0}(\bar z)\propto \phi_{r}^{s=0}(\bar z)\times
\ln(z_+/z_-)+ \text{polynomial in $\bar z$}$
[see Eq.\ (24) in Sec.\ 3.6.2 of \cite{hg}].
Since $\phi_{r}^{s=0}$ is physically regular, it is therefore
clear that $\phi_{ir}^{s=0}$ is physically irregular.

In conclusion, because the function $F_{k=0}^{s=0}$ must be a static
solution physically regular at the EH, it must be proportional
to $\phi_r^{s=0}$. Therefore, Eq.\ (\ref{F0}) applies to the
scalar field too, with $\phi_r$ given in Eq.\ (\ref{P}).

\section{Late time behavior at the EH}\label{EH}

The LTE, Eq.\ (\ref{LTE}), is an expansion in inverse
powers of advanced time $v$, with $r$-dependent coefficients.
Since along the EH itself $r$ is constant and $v$ takes finite
values, this expansion seems especially convenient for analyzing the
``late time'', $v\gg M$, behavior of the fields at the EH.
Potentially, this behavior should be described by the $k=0$ of the
LTE. However, a possible divergence or vanishing of various
``coefficient'' functions $F_k(r)$ at $r=r_+$ may alter this simple
picture, leading to a different prediction for the late time
power-law decay at the EH.
Indeed, as it turns out in this section, there is a case
(the one of $s>0$, $m=0$) in which the term $k=1$ is found to
dominate the term $k=0$ at the EH.

It is therefore important to analyze also the behavior of the
$k\geq 1$ terms at the EH. This task is further motivated by
our wish to verify that the LTE is fully
consistent with regularity at the EH: It will be shown that for
each $k$ there exists a solution $F_k$ physically
regular at the EH. These EH-regular functions will then construct,
via the LTE, a field $\Psi^s$ representing a physical perturbation
which is regular along the EH at all $v$.

With the above motivations in mind, we first derive in this section
expressions for the EH-asymptotic behavior of each of the functions
$F_{k\geq 1}$. The results are summarized in Table \ref{table1}
below. We then discuss the application of these results to the
late time tail along the EH.

\subsection{Behavior of the $k>0$ terms at the EH}
\label{subVIA}

Each of the functions $F_{k>0}^l$ admits the inhomogeneous equation
$D^lF_k^l=Z_k^l$ [Eq.\ (\ref{BasicOrdinary})], and is subject to the
EH-regularity condition (\ref{BC}).
For each $k>0$, the general solution to Eq.\ (\ref{BasicOrdinary})
has the form
\begin{equation} \label{general}
F_{k>0}^{l}(r)=
a_k\phi_r(r) + b_k\phi_{ir}(r) + \phi_k^{ih}(r),
\end{equation}
where $a_k$ and $b_k$ are (yet) arbitrary coefficients,
$\phi_r$ and $\phi_{ir}$ are two independent homogeneous
solutions (those derived in the previous section),
and $\phi_k^{ih}(r)$ is a solution to the inhomogeneous equation.

For each $k>0$, an inhomogeneous solution $\phi_k^{ih}$
is given by
\begin{equation} \label{ih1}
\phi_k^{ih}(r)=
\phi_r(r) \int^r
\frac{\phi_{ir}(r')Z_k(r')/\Delta(r')}{W(r')}\,dr'-
\phi_{ir}(r) \int^r
\frac{\phi_r(r')Z_k(r')/\Delta(r')}{W(r')}\,dr',
\end{equation}
in which
\begin{equation} \label{W}
W=\Delta^{-s-1}
\end{equation}
is the Wronskian associated with the homogeneous equation
$D^lF_k^l=0$.
We can now make use of the relation
\begin{equation} \label{integral}
\phi_{ir}= -\phi_r(r) \int^r \phi_r^{-2}(r')W(r')dr'
\end{equation}
to re-express the above inhomogeneous
solution in a more convenient form, as
\begin{equation} \label{ih2}
\phi_k^{ih}(r)=
\int_{r_1}^r \!\!\!dr' \int_{r_2}^{r'} \!\!\!dr''\,
\frac{\phi_r(r)\phi_r(r'')}{[\phi_r(r')]^2}\,
\frac{W(r')}{W(r'')}\, \frac{Z_k(r'')}{\Delta(r'')},
\end{equation}
where $r_1$ and $r_2$ are constant integration limits.\footnote{
The constant limits $r_1$ and $r_2$ will be specified as convenient
for each of the various cases analyzed below in separate.
Of course, changing these limits amounts to adding a homogeneous
solution to $F_k^l$, which is merely equivalent to re-defining the
coefficients $a_k$ or $b_k$ in Eq.\ (\ref{general})}
This form is obtained from Eq.\ (\ref{ih1}) by first
substituting for $\phi_{ir}$, using Eq.\ (\ref{integral}),
and then integrating the resulting expression by parts.
It is advantageous in that it only involves the
homogeneous solution $\phi_r$, which is of a more simple form
than $\phi_{ir}$ in all the cases considered in the previous
section.

Equation (\ref{ih2}) can be used, in principle, to calculate all
functions $F_k^l(r)$ in an inductive manner.
In general, for each $k>0$ the source function $Z_k^l$
depends on various $l$ modes of the functions $F_{k-1}$ and $F_{k-2}$,
which are to be calculated at previous steps of the induction procedure.
As we show below, for each $k$, the value of one of the coefficients
$a_k$ or $b_k$ is dictated by regularity at the EH.
The other coefficient is to be specified by matching at large distance,
as we explain in Sec.\ \ref{Global}.

We now use Eq.\ (\ref{ih2}) to obtain the EH-asymptotic forms
of all functions $F_{k>0}^l$.
The special case $s>0,\,m=0$ shall be treated separately from all
other cases.

\subsubsection{The case $s\leq 0$ and the case $s>0,\,m\neq 0$}

We start with $k=1$. The source $Z_{k=1}^l$ is calculated from Eq.\
(\ref{Z}), in which, according to the results of the previous
section, we set $F_{k=0}^l=a_0^l\phi_r^l$.
For $s\leq 0$, and also for $s>0$ with $m\neq 0$, we find by Eqs.\
(\ref{basis1}), (\ref{phi0b}), and (\ref{P}) that the function
$F_{k=0}^l$ has the form
$F_{k=0}^l\propto\phi_r^l=\Delta^{-s+im\gamma}\times f_0(r)$,
where $f_0(r)$ is a certain function analytic at the EH
and nonvanishing there.
Substituting this form in Eq.\ (\ref{Z}) yields
\begin{equation} \label{S1}
Z_{k=1}^l(r)=\Delta^{-s+im\gamma}\times \bar f_1(r),
\end{equation}
where $\bar f_1(r)$ is a function analytic at the EH.\footnote{
In deriving Eq.\ (\ref{S1}) one should notice that when $F_{k=0}^l$
is substituted in Eq.\ (\ref{Z}), the term containing $dF_{k=0}^l/dr$
and the one containing $F_{k=0}^l/\Delta$ cancel out at the leading order
in $\Delta$. As a consequence, leading-order contributions to $Z_{k=1}^l$
arise from all terms in (\ref{Z}), including the interaction terms.}

With Eqs.\ (\ref{W}) and (\ref{S1}), Eq.\ (\ref{ih2}) becomes
(for $k=1$)
\begin{equation} \label{int1}
\phi_{k=1}^{ih}=\phi_r(r)\int_{r_+}^{r}dr'\int_{r_2}^{r'}dr''\,
\frac{[\Delta(r'')]^{-s+2im\gamma}}
{[\phi_r(r')]^2 [\Delta(r')]^{s+1}} \times \tilde f_1(r''),
\end{equation}
where $\tilde f_1(r'')$ is analytic at the EH, and
where we have specified the lower limit of the integration over $r'$
as $r_1=r_+$.
Integrating over $r''$ [recalling $\Delta=(r-r_+)(r-r_-)$],
we obtain
\begin{eqnarray} \label{int2}
\phi_{k=1}^{ih}&=&\phi_r(r)\int_{r_+}^{r}dr'\,
\frac{[\Delta(r')]^{-s+1+2im\gamma}+\bar c}
{[\phi_r(r')]^2 [\Delta(r')]^{s+1}} \times \hat f_1(r')
\nonumber\\
&=&\phi_r(r)\int_{r_+}^{r}dr'\,
\left(\tilde c+\bar c\, [\Delta(r')]^{s-1-2im\gamma}\right)
\times \bar{\bar f}_1(r'),
\end{eqnarray}
where $\hat f_1(r')$ and $\bar{\bar f}_1(r')$ are analytic at the EH,
$\tilde c$ is a certain nonvanishing constant, and $\bar c$ is
an integration constant, which, of course, depends on the value
of the lower integration limit $r_2$.
In the case $s\leq 0$, a convenient choice is $r_2=r_+$,
which makes $\bar c$ vanish. For $s>0$ we take $r_2=2 r_+$ (say), as the
choice $r_2=r_+$ is forbidden\footnote{
In the case $s=+2$ we can make $\bar c$ vanish by taking $r_2=\infty$.
However, for $s=+1$ no choice of $r_2$ nullifies $\bar c$.}.
In that case, the contribution proportional to $\bar c$ to the integral
in Eq.\ (\ref{int2}) must coincide (up to a multiplicative constant)
with one of the static solutions,
$\phi_r$ or $\phi_{ir}$, because changing the integration limit $r_2$
(thus changing $\bar c$) amounts to adding a static solution to
Eq.\ (\ref{general}).
By integrating over $r'$ we find that this contribution has the
asymptotic form $\propto \bar c\,\Delta^{-im\gamma}\propto
\bar c\, e^{-im\Omega_+r_*}$ at the EH;
hence it must admit the global form $\propto \bar c\,\phi_{ir}$
[see Eq.\ (\ref{basis2Asy})].
The term proportional to $\bar c$ in $\phi_{k=1}^{ih}$ can therefore be
absorbed in the term $b_{1} \phi_{ir}$ of Eq.\ (\ref{general}), by
re-defining the coefficient $b_{1}$.
One is left with the contribution proportional to $\tilde c$, which,
after integrating over $r'$, reads $\phi_r\cdot\Delta\, f_1(r)$,
where $f_1(r)$ is a function analytic at the EH.

From Eq.\ (\ref{general}) we now find that
$F_{k=1}^l=a_1\phi_r+b_1\phi_{ir}+ \phi_r\Delta f_1(r)$.
The EH-regularity criterion (\ref{BC}) then dictates
$b_1=0$, finally leading to
\begin{eqnarray} \label{F1}
F_{k=1}^l=\phi_r[a_1+ \Delta\cdot f_1(r)].
\end{eqnarray}
Recall (i) that $f_1(r)$ is analytic at the EH, and (ii) that
the regular solution is physically regular there in the sense
discussed in previous sections.
This implies that the function $F^l_{k=1}$ of Eq.\ (\ref{F1}) is
physically regular at the EH (namely, the ``physical'' function
$\hat F_{k=1}^l$ associated with $F_{k=1}^l$ is mathematically
regular there).
Note also that, as far as the leading order term in $\Delta$ is
concerned, $F_{k=1}^l$ has the same asymptotic behavior as
$F_{k=0}^l$ at the EH, which is that of the regular static solution:
$\propto \Delta^{-s}e^{im\Omega_+ r_*}$.

We can now carry on in an inductive manner, and analyze the terms
$k\geq 2$. In general, let us assume that for a given $k'\geq 2$ we
have for all $k<k'$
\begin{eqnarray} \label{Fk}
F_k^l=\phi_r[a_k+ \Delta\cdot f_k(r)],
\end{eqnarray}
where $f_k(r)$ are functions analytic at the EH.
(This form was already verified above for $k=0$ and $k=1$.)
By substituting in Eq.\ (\ref{Z}) it is straightforward to show that
\begin{equation} \label{Sk}
Z_{k'}^l(r)=\Delta^{-s+im\gamma}\times \bar f_{k'}(r),
\end{equation}
where $\bar f_{k'}(r)$ is analytic at the EH.
This, following the same calculation as for $F_{k=1}^l$, leads to
$F_{k'}^l=a_{k'}\phi_r+b_{k'}\phi_{ir}+ \phi_r\Delta f_{k'}(r)$.
The EH-regularity condition (\ref{BC}) then dictates $b_{k'}=0$,
and one finds that Eq.\ (\ref{Fk}) is also valid for $k=k'$.
Thus, by induction, Eq.\ (\ref{Fk}) is verified for all $k\geq 0$.

In conclusion, we have constructed EH-regular solutions for each of
the functions $F_k^l(r)$.
It was found that for all modes $l$, all functions $F_k^l(r)$ behave
near the EH like a regular static solution, $\propto\phi_r$; namely,
they all admit the asymptotic form
\begin{equation} \label{FkAsy}
F_k^l(r)\cong a_k\,\Delta^{-s}e^{im\Omega_+r_*}\quad
\text{near the EH, for all $k\geq 0$}.
\end{equation}

\subsubsection{The case $s>0$, $m=0$}

In this case, the function $F_{k=0}^l=a_0\phi_r^+$ is simply a polynomial,
admitting the EH-asymptotic form $F_{k=0}^{l,m=0}\cong a_0\Delta^0$
[see Eq.\ (\ref{asyphi0EH})].
To obtain the source $Z_{k=1}^l$, insert $F_{k=0}^l$ into Eq.\ (\ref{Z}).
This yields
\begin{equation} \label{S10}
Z_{k=1}^{s>0,l,m=0}=4Msk_0a_0(r_+^2-a^2)\Delta^{-1}+ \bar h(r),
\end{equation}
where $\bar h(r)$ is a function analytic at the EH. Note that now,
since $F_{k=0}^l$ is a polynomial, the asymptotic form of $S_{k=1}^l$
at the EH is dominated by merely the term $\propto \Delta^{-1} F_{k=0}^l$
in Eq.\ (\ref{Z}), while the other terms (including the derivative term
and the interaction terms) contribute only to higher orders in
$\Delta$.

To calculate $\phi_{k=1}^{ih}$, we substitute $S_{k=1}^l$ in Eq.\
(\ref{ih2}). Recalling that in the present case the function $\phi_r$
is a polynomial, one finds
\begin{equation} \label{int3}
\phi_{k=1}^{ih}=4Msk_0a_0(r_+^2-a^2)\phi_r^+(r)\int_{r_1}^{r}dr'
\int_{r_+}^{r'}dr''  \frac{[\Delta(r'')]^{s-1}+[\Delta(r'')]^s
\tilde h(r'')}{[\phi_r^+(r')]^2 [\Delta(r')]^{s+1}},
\end{equation}
where $\tilde h(r'')$ is analytic at the EH. (Here,
we have specified one of the integration limits, $r_2=r_+$.)
Carrying the integration we arrive at
\begin{eqnarray} \label{int4}
\phi_{k=1}^{ih}&=&4Mk_0a_0\,\frac{r_+^2-a^2}{r_+-r_-}\,\phi_r^+(r)
\int_{r_1}^{r}dr' \left[\Delta^{-1}(r')+\hat h_1(r')\right]
\nonumber\\
&=& \tilde\gamma_1\,\phi_r^+\ln z_+ + \bar{\bar h}_1(r),
\end{eqnarray}
in which $\hat h_1(r')$ and $\bar{\bar h}_1(r)$ are analytic at the EH,
and where the coefficient $\tilde\gamma_1$ is given by
\begin{eqnarray} \label{bara}
\tilde\gamma_1= 4Mk_0a_0\,\frac{r_+^2-a^2}{(r_+-r_-)^2}\neq 0.
\end{eqnarray}
By virtue of Eq.\ (\ref{eq45}) we finally obtain for $F_{k=1}^l$ (in
the case $s>0$, $m=0$),
\begin{eqnarray} \label{int5}
F_{k=1}^l(r)&=& a_1^+\phi_r^+ + b_1^+\phi_{ir}^+ +\phi_{k=1}^{ih}
\nonumber\\
&=&a_1^+\phi_r^+ + b_1^+[\Delta^{-s}(1+\tilde\alpha^{+}\Delta+\cdots)
+\tilde\beta\,\phi_r^+\ln z_+] + \tilde\gamma_1\,\phi_r^+\ln z_+\,
+\bar{\bar h}_1(r).
\end{eqnarray}

Now, the EH-regularity criterion (\ref{BC}) forces the ``physical''
function $\hat F_{k=1}^l\equiv \Delta^s F_{k=1}^l$
to be perfectly smooth at the EH (where $z_+=0$). This implies that
$F_{k=1}^l$ must contain no logarithmic terms of the form
$\propto\ln z_+$: If such a logarithmic term
is present, $\hat F_{k=1}^l$ would indeed be continuous at the EH,
yet its $s$'th derivative with respect to $r$ or $U$
(which are regular coordinates at the EH) would diverge there.
We therefore find that the regularity condition dictates the value of the
coefficient $b_1^+$:
\begin{equation} \label{b1}
b_1^+= - \tilde\gamma_1/\tilde\beta\neq 0.
\end{equation}
Hence, from Eq.\ (\ref{int5}) we obtain the form
\begin{equation} \label{F10}
F_{k=1}^{l}(r) =\Delta^{-s} h_1(r),
\end{equation}
where $h_1(r)$ is a function analytic at the EH, satisfying
\begin{equation} \label{g1}
h_1(r=r_+)=b_1^+\neq 0.
\end{equation}
[The analytic function $h_1(r)$ contains also the terms
$a_1^+\phi_r^+$ and $\bar{\bar h}_1(r)$ appearing in Eq.\
(\ref{int5}), multiplied by $\Delta^s$.
Note that the polynomial homogeneous solution does not affect the
leading order term of $F_{k=1}^l$ at the EH.]
It is important that $h_1(r)$ does not vanish at $r=r_+$:
It implies that at the EH itself (in the case of $s>0$ with $m=0$)
the term $F_{k=1}^l$ dominates the term $F_{k=0}^l$, which is only
proportional to $\Delta^0$ there. The application of this result to
the late time tail will be discussed below.

We now turn to the terms $k\geq 2$, and show by mathematical induction
that for all $k\geq 1$ there exists a solution admitting the form
\begin{equation} \label{Fkk}
F_{k\geq 1}^{l}(r) = \Delta^{-s} h_k(r)
\end{equation}
[with $h_k(r)$ being functions analytic at the EH], and thus
satisfying the EH-regularity condition (\ref{BC}).
This form was already verified in the case $k=1$, for which we
also showed that $h_1(r_+)\neq 0$.

Let us assume that Eq.\ (\ref{Fkk}) applies for all $1\leq k< k'$
where $k'>1$ is arbitrary, and show that it is also valid for $k=k'$.
Substituting the form (\ref{Fkk}) into Eq.\ (\ref{Z}) we find
$Z_{k'}^l=\Delta^{-s} \bar h_{k'}(r)$ where $\bar h_{k'}(r)$
is analytic at the EH. We then have
\begin{eqnarray} \label{int6}
\phi_{k'}^{ih}=\int_{r_1}^r dr'\int_{r_+}^{r'} dr''
\frac{\phi_r^+(r)\phi_r^+(r'')}{[\phi_r^+(r')]^2}
\frac{\bar h_{k'}(r'')}{[\Delta(r')]^{s+1}}&=&
\phi_r^+(r)\int_{r_1}^r dr' \Delta^{-s}(r')\tilde h_{k'}(r')\nonumber\\
&=&\tilde\gamma_{k'}\phi_r^+(r)\ln z_+ + \Delta^{-s+1}\hat h_{k'}(r),
\end{eqnarray}
where $\bar h_{k'}$, $\tilde h_{k'}$, and $\hat h_{k'}$ are analytic at
the EH, and $\tilde\gamma_{k'}$ are constant coefficients.
(Here we have taken $r_2=r_+$.) For $F_{k'}^l$ we thus obtain
\begin{equation} \label{int7}
F_{k'}^{l}(r)=
a_{k'}^+\phi_r^+ + b_{k'}^+[\Delta^{-s}(1+\tilde\alpha^{+}\Delta+\cdots)
+\tilde\beta\ln(z)\,\phi_r^+] + \tilde\gamma_{k'}\ln z_+\,\phi_r^+
+\Delta^{-s+1}\hat h_{k'}(r).
\end{equation}
This result is analogous to Eq.\ (\ref{int5}), only here
we have not ruled out the possibility that some of the
coefficients $\tilde\gamma_{k'}$ may vanish (for $k'>1$) .

For the regularity condition to be met, we must now have
$b_{k'}^+=-\tilde\gamma_{k'}/\tilde\beta$.
Consequently, Eq.\ (\ref{Fkk}) is recovered for $k=k'$ as well.
By induction, then, we conclude that Eq.\ (\ref{Fkk}) applies to
all $k\geq 1$.

Recall that it is possible for some of the coefficients
$\tilde\gamma_k$ (with $k\geq 2$) to vanish.
If, for a certain $k$, $\tilde\gamma_k$ happens to vanish,
then, to maintain EH-regularity, one must have $b_k^+=0$.
In that case Eq.\ (\ref{int7}) yields for $F_k^l$ a divergence rate
slower than $\Delta^{-s}$ at the EH (for this specific $k$).
This, of course, does not contradict Eq.\ (\ref{Fkk}), which
should be regarded as merely setting an upper bound to the divergence
rate of the functions $F_{k\geq 2}^l$.
It is only for $F_{k=1}^l$ that we verified the actual asymptotic
behavior $\propto \Delta^{-s}$, by showing $h_1(r_+)\neq 0$.
This information, however, would be sufficient for the late time
analysis at the EH.

The above results, concerning the behavior of the
physically-regular functions $F_k^l(r)$ at the EH, are arranged
in Table I. The table shows the leading-order forms of these
functions for the various cases studied above. Below we use
these results to discuss the late time behavior
of the fields $\Psi^s$ along the EH.


\begin{table}[h]
\centerline{$\begin{array}{||c|c|c|c||}    \hline\hline
{\text{the case}} &
F_{k=0}    & F_{k=1}     & F_{k\geq 2}
\\                                             \hline\hline
\hspace{2mm} am=0,\,\, s>0  &\hspace{5mm}\Delta^{0}\hspace{5mm}
&\hspace{5mm}\Delta^{-s}\hspace{5mm}
&\hspace{5mm}\leq \Delta^{-s}\hspace{5mm}
\\                                             \hline
\hspace{2mm}\text{all other cases}
&\multicolumn{3}{c||}{\Delta^{-s}\,e^{im\Omega_+r_*}
\text{\ for all $k\geq 0$}}
\\                                             \hline\hline
\end{array}$}
\caption{\protect\footnotesize The asymptotic behavior of the
``physically regular'' functions $F_k^l(r)$ at the EH.
Presented are the leading-order in $\Delta$ forms of these
functions, for the various cases studied in the text.
For axially symmetric ($m=0$) modes the asymptotic behavior depends
on whether $s>0$ or $s\leq 0$, as discussed in the text.
Note that these asymptotic forms are in all cases independent of the
multipole number $l$ of the modes under consideration.
}\label{table1}
\end{table}

\subsection{The late time tail along the EH}
\label{subVIB}

When discussing the field behavior along the EH, it is most natural
to refer to the ``physical field'' $\hat\Psi^s=\Delta^s\Psi^s$,
which, by construction, is a linear combination (with regular coefficients)
of the regular Weyl or Maxwell components.
Expressing the LTE in terms of $\hat\Psi^s$, and using Eq.\
(\ref{phi_irregular}), we find
\begin{equation} \label{EH1}
\hat \Psi^s(v,r,\theta,\tilde\varphi_+)=
\sum_{k=0}^{\infty} \sum_{l,m}\,
Y^{slm}(\theta,\tilde\varphi_+)\, e^{im\Omega_+v}
\left[\Delta^s e^{-im\Omega_+r_*} F_k^{slm}(r)\right]
\, v^{-k_{0}-k}.
\end{equation}
Here, the factor in the square brackets is the function $\hat F_k^l(r)$
which, by the above construction, is regular at the EH (for all $k$).
Recall also that the angular dependence here is EH-regular, and that the
$v$-dependent factors take finite values at the EH. Thus,
each of the terms in the sum over $k$ in Eq.\ (\ref{EH1}) is indeed
physically regular at the EH.

Now, at large $v$, the field $\Psi^s$ should, potentially, be
dominated by the $k=0$ term in Eq.\ (\ref{EH1}).
For the $s\leq 0$ fields and for $m\neq 0$ modes of the $s>0$ fields,
we find from Table \ref{table1} that at the EH itself
the factor in the square brackets in Eq.\ (\ref{EH1}) admits
$\left[\ \right]_{k}\propto {\rm const}$ for all $k\geq 0$.
Therefore, in these cases, the late time decay of $\Psi^s$ along
the EH is indeed dominated by the $k=0$ term, with other terms
smaller by factors of $1/v$.
Let us denote by $\Psi^{sm}$ the part of $\Psi^s$ which includes
all multipole modes $l$ of a given $m$.
To leading order in $1/v$, we then find for all modes $m$ of the
$s\leq 0$ fields, and for non-axially symmetric ($m\neq 0$) modes
of the $s>0$ fields,
\begin{equation} \label{EH2}
\hat \Psi^{sm}(r=r_+)= \sum_{l}a_0^l\, Y^{slm}(\theta,\tilde\varphi_+)\,
e^{im\Omega_+v} \, v^{-k_{0}}.
\end{equation}

The situation is different in the case of axially symmetric ($m=0$)
modes of the $s>0$ fields. Here, the
factor in the square brackets in Eq.\ (\ref{EH1}) vanishes at
the EH like $\Delta^s$ for $k=0$, whereas for $k=1$ it is finite.
Hence, in this case, the $k=1$ term dominates the $k=0$ term.
For each of the $k\geq 2$ terms the above factor is at most
finite (and may even vanish for some $k$); hence these terms
are negligible with respect to the $k=1$ term at large $v$.
We conclude that for axially symmetric modes of
the $s>0$ fields the late time behavior along
the EH is dominated by the $k=1$ term:
\begin{equation} \label{EH3}
\hat \Psi^{s>0,m=0}(r=r_+)=\sum_{l}b_1^+ \, Y^{sl,m=0}(\theta)\,
\, v^{-k_{0}+1},
\end{equation}
to leading order in $1/v$
(here, the coefficient $b_1^+$ is also $l$ dependent).

At this stage, we still do not know the mode composition of the
above leading-order tails; neither can we tell what the power
index $k_0$ and the amplitude coefficients are.
These pieces of information will be obtained below, by
matching the LTE to the form of the late time field at null
infinity.

However, one feature of the behavior along the EH is already
manifested in Eq.\ (\ref{EH2}) above:
Non-axially symmetric ($m\neq 0$) modes of the fields do not exhibit
a strict power-law decay along the EH;
rather, the amplitude of the power-law tail {\em oscillates} along the
null generators of the EH, with an (advanced time) frequency $m\Omega_+$.
This phenomenon was first observed by Ori \cite{AmosI}, and was
further analyzed in Refs.\ \cite{Letter,HodI,HodII}.

We comment that the above oscillations are not manifested when
using the Kerr coordinate $\tilde\varphi$, defined by
$d\tilde\varphi=d\varphi +(a/\Delta)dr$, instead of
$\tilde\varphi_+$ (cf.\ Ref.\ \cite{KrivanI}, which adopts the
coordinate $\tilde\varphi$). Both coordinates are regular at the
EH; however, the horizon's null generators are lines of constant
$\tilde{\varphi}_{+}$ but varying $\tilde{\varphi}$.\footnote{ To
see that, we point out that, in the $(v,r,\theta,\tilde\varphi)$
system, the null generators of the EH are lines of
$\theta={\rm const}$, $r=r_+$, $\tilde\varphi=2a\lambda$, and
$v=2(r_+^2+a^2)\lambda$, where $\lambda$ is an affine parameter
along the generators---see \S33.6 in Ref.\ \cite{MTW}.
Thus, along the null generators we find $\tilde\varphi=\Omega_+ v$.
Now, at the EH the two coordinates $\tilde\varphi$ and
$\tilde\varphi_+$ are related by $d\tilde\varphi=d\tilde\varphi_+
+ \Omega_+dv$, from which we conclude that $d\tilde\varphi_+=0$ along
the null generators of the EH.}
Note that the oscillation of the scalar field along the horizon's
null generators is a coordinate-independent phenomenon.

\section{Late time behavior at fixed $\lowercase{r>r_+}$}
\label{Global}

In this section we obtain the global late time behavior
of any of the modes of the field $\Psi^{s}$ at any fixed value of
$r>r_+$.
This task is to be accomplished in two steps:
First, we consider the mode $l=l_0$ (for each given $s$ and $m$),
which in Sec.\ \ref{NullInfinity} we found to be the (single)
dominant mode at null infinity at late time.
By evaluating the form of this mode of the LTE
at null infinity, and comparing it to the form obtained independently
in Sec.\ \ref{NullInfinity}, we derive the unknown power index $k_0$,
as well as the amplitude coefficient of the late time tail at fixed $r$.
Provided with the value of $k_0$, we can then carry on and,
in the second step, obtain the tail form of all other modes at fixed
$r>r_+$. In particular, we then find that the single mode $l_0$
dominates the behavior of the field $\Psi^{sm}$ also at fixed $r$.

\subsection{The mode $l=l_0$}
\label{subVIIA}

Substituting $v=t+r_*$ in Eq.\ (\ref{LTE}), we find that along any
$r={\rm const}>r_+$ world line, at the late time limit $t\gg |r_*|$, the
behavior of the mode $\Psi^{sl_0m}$ (for each given $s$ and $m$)
is described by
\begin{eqnarray} \label{global0}
\Psi^{sl_0m}&=&  Y^{sl_0m}(\theta,\varphi)\,F_{k=0}^{sl_0m}(r)\,t^{-k_{0}}
\nonumber\\
&=&a_0^{l_0}\,Y^{sl_0m}(\theta,\varphi)\,\phi_r^{l_0}(r)\,t^{-k_{0}}
\end{eqnarray}
to leading order in $|r_*|/t$, where the second equality is due
to Eq.\ (\ref{F0}).
To obtain the unknown power index $k_0$ and the amplitude coefficient
$a_0^{l_0}$, we now match the LTE to the form of the field at null
infinity, as derived in Sec.\ \ref{NullInfinity}.

In order for Eqs.\ (\ref{Psil0}) and (\ref{LTE})
to agree at null infinity for the mode $l=l_0$, we must have
\begin{equation} \label{matching}
\left[\sum_{k=0}^{\infty}\,F_k^{sl_0m}(r)\,v^{-k_{0}-k}\right]_
{\text{at scri+}} \equiv \psi_{\rm LTE}^{l_0}=
\alpha_{0}^{l_0}\,I_0^{l_0}\, v^{-2s-1}\,u^{-(l_0-s+2)}.
\end{equation}
Here, $\psi_{\rm LTE}^{l_0}$ denotes the time-radial
part of the mode $l_0$ at null infinity, as calculated from the LTE,
whereas the expression on the right-hand side (RHS) is the one derived
in Sec.\ \ref{NullInfinity} using the iterative expansion scheme.
As it turns out (see below), all terms $k$ of the sum on the LHS of
this equation contribute in the same order of magnitude at null
infinity, and thus should be all summed up when evaluating
$\psi_{\rm LTE}^{l_0}$. To that end, we first need to obtain the
large-$r$ asymptotic form of all functions $F_k^{l_0}(r)$.

Starting with $k=0$, we have from Eq.\ (\ref{F0}),
$F_0^{sl_0m}=a_0^{l_0}\phi_r^{l_0}$.
Using Eqs.\ (\ref{basis1Asy}) and (\ref{asyphi0INF}) we find the
large-$r$ asymptotic form
\begin{equation} \label{infinity1}
F_{k=0}^{sl_0m}(r\gg r_+)\cong \gamma_0^{l_0}\,r^{l_0-s},
\end{equation}
where the constant coefficient $\gamma_0^{l_0}$ is given by
\begin{equation} \label{gamma0}
\gamma_0^{l_0}= a_0^{l_0}\times\left\{
\begin{array}{ll}
B_{-s}^{l_0,m=0} & \text{for $s>0$ with $m=0$},\\
B_{s}^{l_0m}  & \text{in all other cases}
\end{array}
\right.
\end{equation}
[with the coefficients $B_{s}^{l_0m}$ given in Eq.\ (\ref{B})].

To analyze the functions $F_{k\geq 1}^{l_0}$ we use Eq.\
(\ref{general}), with the coefficients $b_k$ taken to be the ones
determined above by EH-regularity considerations (e.g.\ $b_k=0$ for
$m\neq 0$).
We now show by mathematical induction that for all $k$,
these functions admit the large-$r$ asymptotic form
\begin{equation} \label{infinity2}
F_{k}^{l_0}(r\gg r_+)\cong \gamma_k^{l_0}\, r^{l_0-s+k}
\end{equation}
(to leading order in $r$), where $\gamma_k^{l_0}$ are constant
coefficients. To that end, we assume that Eq.\ (\ref{infinity2})
applies to all $k<k'$ (where $k'$ is an arbitrary integer
greater than zero), and verify its validity to $k=k'$.

First, we must calculate the large-$r$ asymptotic form of the
source function $Z^{l_0}_{k'}$. From Eq.\ (\ref{Z}) we obtain, to
leading order in $r$,
\begin{equation} \label{infinity3}
Z_{k'}^{l_0}\cong
2(k_0+k'-1)\left[r^2(dF_{k'-1}^{l_0}/dr)+rF_{k'-1}^{l_0}
\right] \cong \hat\gamma_{k'}^{l_0}\, r^{l_0-s+k'},
\end{equation}
where $\hat\gamma_{k'}^{l_0}=2(k_0+k'-1)(l_0-s+k')\gamma_{k'-1}^{l_0}$
are constants.\footnote{
Actually, the source $Z_{k'}^{l_0}$ contains also contributions
from other modes ($l=l_0+1,l_0+2$). However, as we show later in this
section, such contributions are negligible at large $r$, and do not
affect the asymptotic form (\ref{infinity3}).}

Substituting this leading-order form in Eq.\ (\ref{ih2}) and
performing the double integration, we obtain, to leading order in $r$,
$\phi_{k'}^{ih}\cong \gamma_{k'}^{l_0}r^{l_0-s+k'}$,
with
\[
\gamma_{k'}^{l_0}=\frac{\hat\gamma_{k'}^{l_0}}{k'(2l_0+k'+1)}=
\frac{2(k_0+k'-1)(l_0-s+k')}{k'(2l_0+k'+1)}\,\gamma_{k'-1}^{l_0}.
\]
Thus, for all $k'>0$, the contribution from $\phi_{k'}^{ih}$ to
$F_{k'}^{l_0}$ at large $r$ [via Eq.\ (\ref{general})] dominates the
contribution from the homogeneous solutions, which is at most
$\sim r^{l_0-s}$. We therefore have
$F_{k'}^{l_0}\cong \phi_{k'}^{ih}$ at large $r$; hence
Eq.\ (\ref{infinity2}) is satisfied for $k=k'$ as well.
We also know by Eq.\ (\ref{infinity1}) that Eq.\ (\ref{infinity2})
is valid for $k=0$.
This, by mathematical induction, verifies Eq.\ (\ref{infinity2})
for all $k\geq 0$, with the coefficients $\gamma_k^{l_0}$ given by
\begin{equation} \label{gammak}
\gamma_k^{l_0}=\frac{(2l_0+1)!\,\gamma_0^{l_0}}{(k_0-1)! (l_0-s)!}\,\,
\left[\frac{2^k(k_0+k-1)!(l_0-s+k)!}{k!(2l_0+k+1)!}\right].
\end{equation}

We are now in position to evaluate the sum over $k$ on the LHS
of Eq.\ (\ref{matching}) at null infinity.
Substituting for $F_k(r)$ [using Eq.\ (\ref{infinity2})], and
recalling $r\cong r_*=(v-u)/2$ at large $r$, we obtain
\begin{equation} \label{sum1}
\psi^{l_0}_{\rm LTE} \cong v^{l_0-s-k_0} \sum_{k=0}^{\infty}\gamma_k^{l_0}
\left[\frac{1}{2}\left(1-\frac{u}{v}\right)\right]^{l_0-s+k}.
\end{equation}
To evaluate the sum of this power series at null infinity,
we make use of the auxiliary identity, valid for $|q|<1$,
\begin{equation} \label{sum2}
\sum_{k=0}^{\infty}\frac{(k_0+k-1)!(l-s+k)!}{k!(2l+k+1)!}\,q^k=
q^{-2l-1}\left[q^{k_0-1}\left(\frac{q^{l-s}}{1-q}\right)^{(l-s)}\right]
^{(k_0-2l-2)},
\end{equation}
where the derivatives on the RHS are with respect to $q$.
[To prove this identity, insert the power expansion
$\frac{q^{l-s}}{1-q}=q^{l-s}\sum_{k=0}^{\infty}q^k$ into the RHS.]
We now make the substitutions $q\to (1-u/v)$ and $l\to l_0$.
At $v\gg u$ (recall that at null infinity $v\to \infty$
whereas $u$ takes finite values), the expression on the RHS
of Eq.\ (\ref{sum2}) is then dominated by the term resulting from
$k_0-l-s-2$ differentiations of the factor $(1-q)^{-1}$ ---
which yields $(k_0-l_0-s-2)!(u/v)^{-(k_0-l_0-s-1)}$.
Substituting this result in Eq.\ (\ref{sum1}), we obtain
\begin{equation} \label{sum3}
\psi^{l_0}_{\rm LTE} \cong
\frac{2^{-l_0+s}(2l_0+1)!(k_0-l_0-s-2)!\,\gamma_0^{l_0}}{(k_0-1)!(l_0-s)!}
\,v^{-2s-1}\,u^{-(k_0-l_0-s-1)}.
\end{equation}

Comparison of Eqs.\ (\ref{matching}) and (\ref{sum3}) finally yields
\begin{equation} \label{k0}
k_0=2l_0+3,
\end{equation}
and also, with the help of Eq.\ (\ref{gamma0}),
\begin{equation} \label{coefficient}
a_0^{l_0}=\alpha_0^{l_0}I_0^{l_0}\times
\frac{2^{l_0-s+1}(l_0+1)}{l_0-s+1}\times
\left\{\begin{array}{ll}
1/B_{-s}^{l_0,m=0} & \text{for $s>0$ with $m=0$},\\
1/B_{s}^{l_0m}  & \text{in all other cases}.
\end{array}\right.
\end{equation}

The parameter $k_0$ derived above is, in view of Eq.\ (\ref{global0}),
the power index of the $l_0$ mode's tail at fixed $r>r_+$ for each
given $s$ and $m$ [recall $l_0=\max(|s|,|m|)$].
The coefficient $a_0^{l_0}$ describes the amplitude of this tail,
with Eq.\ (\ref{coefficient}) relating it to the amplitude of
the $l_0$-mode's tail at null infinity
(which is $\alpha_0^{l_0} I_0^{l_0}$).

\subsection{The modes $l>l_0$}
\label{subVIIB}

We now turn to analyze the behavior of the other modes, $l>l_0$,
at fixed $r$. Here, the coupling between modes in the Kerr
case shall appear to have a crucial effect on the form of the
late time tail.
To discuss this effect, it is most instructive to first
consider a situation without coupling.
Thus, at first, we shall ``turn off'' the interactions between
modes by ignoring for a while all terms in $Z_k^l$ [Eq.\ (\ref{Z})]
which couples the mode $l$ to other modes.
(This will qualitatively describe the situation in the Schwarzschild
case.)
Then, in the second part of the following discussion, we restore the
coupling (by taking into account all terms in $Z_k^l$),
and discuss its important effect on the late time tail of decay.

\subsubsection{The case with no coupling between modes}

Considering a mode $l>l_0$ and ignoring its interaction
with other modes,
we can follow the same calculation as for the $l_0$ mode,
and obtain at null infinity
$\psi_{LTE}^l\propto a_0^l\,v^{-2s-1}u^{-(k_0-l-s-1)}$,
in a full analogy with Eq.\ (\ref{sum3}).\footnote{
Recall that in obtaining Eq.\ (\ref{sum3}) for the $l_0$ mode,
the interaction between
modes was not taken into account (this will be justified below).
Thus, the analogy with this case is straightforward: one only needs
to replace $l_0$ with $l$ in Eq.\ (\ref{sum3}).
}
However, since for any $l>l_0$ we have $k_0-l-s-1=2l_0-l-s+2<l-s+2$,
this result cannot match the known $u$ power at null infinity,
$u^{-(l-s+2)}$, unless $a_0^l=0$. The ``boundary condition'' at null
infinity thus dictates the vanishing of $a_0^l$, and thus of $F_0^l$,
for all $l>l_0$. This, of course, means that the modes $l>l_0$ will decay
faster than $t^{-k_0}$ at fixed $r$.

This argument demonstrates that in order to
determine the form of the late time tail at fixed $r$, it is necessary
to find out, for each given mode $l>l_0$, what is the smallest $k$ value
for which $F_k^l$ does not vanish.
Denoting this value by $\tilde k(l)$, we find that the tail of
any mode $l$ is given by $t^{-k_0-\tilde k(l)}=t^{-2l_0-3-\tilde k(l)}$.

What is $\tilde k(l)$ then? We shall answer this question by
matching the LTE at null infinity for each mode $l>l_0$.
First, we must obtain the asymptotic form of the functions $F_k^l$
at large $r$.
By definition of $\tilde k$ we have $F_{k<\tilde k}^l=0$.
If no interaction occurs between various modes, then
the function $F_{k=\tilde k}^l$ must be a static solution
(the regular one):
$F_{k=\tilde k}^l=a_{\tilde k}^l\phi_r\cong a_{\tilde k}^l r^{l-s}$
at large $r$, where the constant coefficient $a_{\tilde k}^l$ does
not vanish by definition of $\tilde k$.
Following the same calculation as for the $l_0$ mode (again, with the
interaction between modes ignored for a while), we obtain for each
$k\geq \tilde k$ the large-$r$ asymptotic form
\begin{equation} \label{FkAsy2}
F_{k\geq\tilde k}^l\cong \gamma_k^l\, r^{l-s+(k-\tilde k)},
\end{equation}
where $\gamma_k^l$ are constant coefficients.
Hence, for each mode $l>l_0$ we find, to leading order in $u/v$,
\begin{equation} \label{sum4}
\psi_{\rm LTE}^l \cong
\sum_{k=\tilde k}^{\infty}\gamma_k^l\, r_{*}^{l-s+k-\tilde k}
\,v^{-k_0-k} = v^{-k_0+l-s-\tilde k}\times f(u/v),
\end{equation}
where $f\equiv \sum_{k=\tilde k}^{\infty}\gamma_k^l
[(1-u/v)/2]^{l-s+k-\tilde k}$ is a function of $u/v$ only.\footnote{
Here we do not sum over $k$ in an explicit manner, as we did for
the $l_0$ mode. Rather, we use a simpler argument, which is yet
somewhat less rigorous.}

Now, we know independently from Sec.\ \ref{NullInfinity} that any
$l$ mode decays at
null infinity with a tail of the form $\propto v^{-2s-1}u^{-(l-s+2)}$.
By comparing the $v$ power in Eq.\ (\ref{sum4}) to this form, we
find (setting $k_0=2l_0+3$) that the function $f(u/v)$ must admit the
asymptotic form $f(u/v)\cong (u/v)^{-2l_0-\tilde k+l+s-2}$.
Then, by comparing the power of $u$, we finally obtain
\begin{equation} \label{tildek}
\tilde k(l) =2(l-l_0)\quad \text{(no coupling; SBH case)}.
\end{equation}

We conclude that for each mode $l>l_0$, the form of the field at
null infinity (acting as a boundary condition) dictates the vanishing
of all term of the LTE with $k<2(l-l_0)$.
The first nonvanishing term, the one with $k=2(l-l_0)$, exhibits
the late time tail $t^{-k_0-2(l-l_0)}=t^{-2l-3}$.
Other terms, the ones with $k>2(l-l)$, decay faster at late time.
Therefore, substituting $v=t-r_*$ in Eq.\ (\ref{LTE}),
we find for each of the modes $l\geq l_0$,
\begin{equation} \label{tailSch}
\Psi^{slm}(t\gg |r_*|)\propto
Y^{slm}(\theta,\varphi)\,\phi_r^l(r)
\, t^{-2l-3}\quad \text{(no coupling; SBH case)}
\end{equation}
to leading order in $|r_*|/t$.

\subsubsection{The effect of coupling between modes}

Let us now ``turn back on'' the interactions between modes, and
consider their effect on the late time tail.
For each of the functions $F_k^l$, the inhomogeneous part
$\phi_{ih}^l$ in Eq.\ (\ref{general}) now contains contributions
not only from the functions $F_{k-1}^l$ and $F_{k-2}^l$ but also
from $F_{k-2}^{l\pm 2}$, and (for $s\neq 0$) from $F_{k-1}^{l\pm 1}$
and $F_{k-2}^{l\pm 1}$.
For example, the function $F_{k=1}^{l_0+1}$ admits (for $s\neq 0$)
a nonvanishing source $\propto aF_{k=0}^{l_0}= aa_{0}^{l_0}\phi_r^{l_0}$.
Since we have $a_0^{l_0}\neq 0$ (by definition of $k_0$),
then, necessarily, the function $F_{k=1}^{l_0+1}$ is nonvanishing.
Thus, for the mode $l=l_0+1$ we find $\tilde k=1$, implying a late time
tail of the form $t^{-k_0-\tilde k}=t^{-2l_0-4}$.
This is different than in the ``no-coupling'' situation,
in which for the $l=l_0+1$ mode we had $\tilde k=2$
[Eq.\ (\ref{tildek})], leading to the tail $t^{-2l_0-5}$
[Eq.\ (\ref{tailSch})].
We may summarize the result in this example by saying that
the interaction ``excites'' the mode $l_0+1$ already at $k=1$,
whereas the boundary condition at null infinity ``excites'' this
mode only at $k=2$.
We arrive at the conclusion that for this mode the effect
of interaction dominates the late time behavior!

The generalization of this result to all $l>l_0$ is straightforward:
For each $l\geq l_0$, the function $F_{\tilde k}^l$
[recall that $\tilde k(l)$ is, for a given mode $l$, the smallest
$k$ for which $F_k^l$ is nonvanishing]
serves as a source, via $Z_{\tilde k}^l$, to $F_{\tilde k+2}^{l+2}$
and (for $s\neq 0$) to $F_{\tilde k+1}^{l+1}$ and $F_{\tilde k+2}^{l+1}$.
These three functions are then necessarily nonvanishing.
Since we have $\tilde k(l=l_0)=0$, we obtain in the case
$s\neq 0$, $\tilde k(l)=l-l_0$ for all $l\geq l_0$.
In the scalar field case ($s=0$) the interaction couples only between
next-to-nearest modes. In this case, the mode $l=l_0+1$ (for example)
is not excited by interaction, but rather by the boundary condition
at null infinity,
yielding $\tilde k(l=l_0+1)=2$ [see Eq.\ (\ref{tildek})].
In general, for $s=0$ we thus find $\tilde k=l-l_0$ for even $l-l_0$,
and $\tilde k=l-l_0+1$ for odd $l-l_0$.
We can express the results in all the above cases by writing
\begin{equation} \label{tildek2}
\tilde k(l) =l-l_0+q\quad \text{(with coupling}),
\end{equation}
where
\begin{equation} \label{q}
q=\left\{\begin{array}{ll}
1 & \text{for $s=0$ with odd $l-l_0$},\\
0 & \text{otherwise}.
\end{array}
\right.
\end{equation}

By comparing Eqs.\ (\ref{tildek}) and (\ref{tildek2}) we find that
interaction excites any of the modes $l>l_0$ at {\em smaller} $k$
than do the boundary conditions at null infinity.
In other words, it is the {\em interaction} (rather than the boundary
conditions at null infinity) which first excites any of these modes,
and thus determines the form of the leading order tail
at late time. This tail shall admit the form
$t^{-k_0-\tilde k}=t^{-l-l_0-3-q}$.

More precisely, for each given $s$ and $m$, and for each of the modes
$l\geq l_0$, we find from Eq.\ (\ref{LTE}), to leading order in
$|r_*|/t$,
\begin{equation} \label{tailKerr}
\Psi^{slm}(t\gg |r_*|)\propto
Y^{slm}(\theta,\varphi)\,F_{k=l-l_0+q}^{l}(r)
\, t^{-(l+l_0+3+q)}\quad \text{(Kerr case)}.
\end{equation}
Note that for each $l>l_0$ (for the scalar field---for each $l>l_0+1$)
the late time tail decays {\em slower} than the corresponding tail of
the same mode in the Schwarzschild case, Eq.\ (\ref{tailSch}).
These slowly decaying tails are produced by interaction, as discussed
above.
On the other hand, the mode $l_0$ (for each $s$ and $m$), whose form
is not shaped by interaction, exhibits the same decay rate in both the
SBH and the KBH cases: $t^{-2l_0-3}$.

There now remains an important question to deal with:
In what way does the matching of the LTE at null infinity change under
the effect of interaction between modes? Does the LTE remain consistent
with the boundary condition there?
In particular, we would like to show that the calculation made
above for the $l_0$ mode is still valid even when the interaction is
taken into account.

To answer these questions we must first examine the large-$r$
asymptotic form of the various functions $F_k^l$, this time taking into
account also the effect of interaction between modes.
As we recall from Eq.\ (\ref{general}), each function $F_k^l$ contains
a ``homogeneous'' part, $a_k^l\phi_r^l+b_k^l\phi^l_{ir}$,
and an ``inhomogeneous'' part, $[\phi_k^l]^{ih}$.
The homogeneous part vanishes identically for all $k<2(l-l_0)$
[by virtue of Eq.\ (\ref{tildek})], and behaves as $r^{l-s}$
at large $r$ for $k\geq 2(l-l_0)$.
The inhomogeneous part vanishes identically for all $k<l-l_0+q$
[by virtue of Eq.\ (\ref{tildek2})]. Its large-$r$ form can be
calculated for each $k$ and $l$ using Eqs.\ (\ref{Z}) and (\ref{ih2}),
given the large-$r$ form of all functions $F_{k'}^{l'}$
which serve as sources to $F_k^{l}$.

Using Eqs.\ (\ref{Z}) and (\ref{ih2}), one may formulate two practical
calculation rules:

(i) If the function $F_{k-1}^l$ admits the large-$r$
form  $\sim r^p$ (where $p$ is some positive power index, and a `$\sim$'
symbol represents the asymptotic form to leading order in $r$), then
the contribution to $F_k^l$ due to the term in $Z_k^l$
involving $F_{k-1}^l$ would be of order $\sim r^{p+1}$.
For example, the source function $F_{k=0}^{l_0}$ (which is an EH-regular
homogeneous solution and thus admits $\sim  r^{l_0-s}$)
induces on $F_{k=1}^{l_0}$ (through $Z_{k=1}^{l=l_0}$) a nonvanishing
contribution of order $\sim r^{l_0-s+1}$.

(ii) The contribution to a function $F_k^l$ due to interaction
with a source function $F_{k'}^{l'\neq l}$ admits the same
large-$r$ form as the source function itself.
This is the situation with all interaction sources of $F_k^l$
and, also, with the source function $F_{k-2}^{l}$.
For example, the source $F_{k=0}^{l_0}\sim r^{l_0-s}$ induces on
$F_{k=2}^{l_0+2}$ (through $Z_{k=2}^{l=l_0+2}$) a nonvanishing
contribution of order $\sim r^{l_0-s}$.
It also, for example, contributes in order $\sim r^{l_0-s}$
to $F_{k=2}^{l_0}$ (through $Z_{k=2}^{l=l_0}$).

With the above two ``rules of thumb'' one can now inductively
construct expressions for the large-$r$ form of each of the
functions $F_k^l$, starting with $F_{k=0}^{l_0}\sim r^{l_0-s}$.
This easily yields $F_k^l\sim r^{(l_0-s)+k-(l-l_0)}$, namely,
\begin{equation} \label{FkAsy3}
F_k^l(r\gg r+)\sim r^{2l_0-l-s+k}
\end{equation}
for each $l\geq l_0$ and $k\geq \tilde k(l)=l-l_0+q$.\footnote{
In fact, this result, Eq.\ (\ref{FkAsy3}), is not completely accurate,
as the functions
$F_{k>\tilde k}^{l>l_0}$ turn out to involve logarithmic factors
which complicate the situation. This logarithmic dependence will
be discussed at the end of this section; meanwhile we shall ignore
it to make the discussion more clear, and refer only to the
power-law dependence of the functions $F_k^l$ (which is not affected
by the presence of the logarithmic factors).}
This is the same asymptotic form as obtained for the functions
$F_k^l$ in the absence of interaction, for $k\geq 2(l-l_0)$ ---
see Eq.\ (\ref{FkAsy2}) with $\tilde k=2(l-l_0)$.
[The important difference is that without interaction
the terms $l-l_0+q\leq k<2(l-l_0)$ all vanish identically.

We conclude the following:

(I) For any mode $l$, the ``inhomogeneous'' part ($\sim r^{2l_0-l-s+k}$)
of $F_k^l$ dominates its ``homogeneous part'' ($\sim r^{l-s}$) at
large $r$ for all $k$ except $k=2(l-l_0)$, where both parts contribute
to order $\sim r^{l-s}$ [recall that the homogeneous contribution
vanishes identically for $k<2(l-l_0)$].
Thus, for each mode $l$, only one arbitrary parameter [the one belonging
to the homogeneous solution at $k=2(l-l_0)$] is involved in the leading
order form at large $r$---as we may expect.
(For each $l$, this parameter is to be determined, in principle, by
matching at null infinity.)
Note also that the first non-vanishing function of each $l$
(namely $F_{\tilde k}^l$), is always proportional to the parameter
$a_0^{l_0}$, which originates from the mode $l=l_0$ at $k=0$, and
``propagates'' through interaction to higher modes.
[This parameter was determined above, Eq.\ (\ref{coefficient}),
by matching the mode $l_0$ at null infinity.]

(II) Whether the interactions are taken into account or not,
one obtains the large-$r$ form $F_k^l\propto r^{2l_0-l-s+k}$
[for all $k\geq 2(l-l_0)$], though with different proportion
coefficients. That difference only affects the amplitude of the
function $\psi_{\rm LTE}^l$ [e.g.\ in Eq.\ (\ref{sum4})],
yet the matching of the modes $l>l_0$ at null infinity, discussed
above, remains qualitatively the same.

(III) As to the effect of the interaction on the mode $l_0$:
By Eq.\ (\ref{FkAsy3}) we have for all $k$, at the leading order in $r$,
$F_k^{l_0}\sim r^{l_0-s+k}$.
This function has, in general, interaction-induced contributions coming
from $F_{k-1}^{l+1}$, $F_{k-2}^{l+1}$, and $F_{k-2}^{l+2}$, which by
Eq.\ (\ref{FkAsy3}) and
the above rule (ii) are of order $\sim r^{l_0-s+k-2}$, $\sim
r^{l_0-s+k-3}$, and $\sim r^{l_0-s+k-4}$, respectively.
Thus, the large-$r$ leading order behavior of the functions
$F_k^{l_0}$ is not affected by interaction with other modes.
This result is valid only for the $l_0$ mode
(and, for $s=0$, also for $l_0+1$) which only interacts with modes of
larger $l$. It justifies ignoring the interactions
when evaluating the behavior of the mode $l_0$ at null infinity,
as we did above.
Particularly, the values derived above for the parameter $k_0$
[Eq.\ (\ref{k0})] and for the coefficient $a_0^{l_0}$
[Eq.\ (\ref{coefficient})] remain valid also with interaction
between modes taken into account.

Finally, it should be commented on that in the above discussion
(regarding the $l>l_0$ modes) we have ignored a certain complication
for the sake of clarity:
Actually, the integration in Eq.\ (\ref{ih2}) produces,
for certain values of $k$ and $l$, a logarithmic dependence on $r$.
This leads to an asymptotic form of $F_k^l$ which is not strictly
a power law [as in Eq.\ (\ref{FkAsy3})], but, in fact, having the form
$F_k^l\sim r^{2l_0-l-s+k}\times (\ln r)^L$. It can be shown
that for each $l$ and $k$, the logarithmic power $L$ may take only
integer values between $0$ and $l-l_0$.
Particularly, we find no logarithmic dependence in all functions
$F_k$ belonging to the mode $l_0$, and thus no modification is required
in the above analysis for this mode.
Also, we can show that $L=0$ for all modes $F_{\tilde k}^l$ (i.e.\ for
the first nonvanishing function $F_k$ of each mode $l$), and therefore
the functions $F_k^l$ in Eq.\ (\ref{tailKerr}) exhibit no logarithmic
dependence at large $r$.
Finally, the matching of the $l>l_0$ modes at null infinity, discussed
above, was based merely on the power-law dependence of the functions
$F_k^l(r)$, which is not affected by the presence of the logarithmic
factors. However, it is not clear to us whether the logarithmic factors
themselves properly match at null infinity [recall that in Sec.\
\ref{NullInfinity} the logarithmic dependence (in $u$) of the modes
$l>l_0$ at null infinity has not been fully investigated].
This question remains open.

\section{summary and discussion}\label{conclusions}

In this paper we have explored analytically
the late time decay of the Newman-Penrose scalars $\Psi^s$
(representing scalar, electromagnetic, and gravitational perturbations)
in the background of a realistic Kerr black hole.
Our analytic method can be summarized as follows:
We assume that at late time each of the fields $\Psi^s$ admits
the {\em late time expansion}, Eq.\ (\ref{LTE}).
This reduces the master perturbation equation to a hierarchy
of ordinary differential equations for the radial functions $F_k^l(r)$.
The homogeneous part of each of these equations is just the static field
equation in Kerr spacetime, to which there exists an analytic basis of
exact solutions. In addition, for each $l$ and $k>0$, each of these
equations possesses an inhomogeneous part depending on functions
$F_{k'<k}$ (including functions which belong to other modes).
Using the Wronskian method we can then explore, in an inductive manner,
the general solution for each of the functions $F_k^l$.
Each of these solutions contains, in advance, two unknown parameters.
One of these parameters is determined by regularity requirements at the
EH. The other parameter is determined by the form of the field at null
infinity (serving as a boundary condition).
To obtain the behavior of the fields at null infinity,
we apply the {\em iterative scheme}, as described in Sec.\
\ref{NullInfinity}.

Following is a summary of our main results.
These results are valid in the most realistic initial setup
of a compact pulse composed of all multipole modes
(and in particular, the lowest radiatable mode $l_0$ for each value
of $m$; below we also briefly discuss the more special case where
this mode is missing).

\subsubsection* {Tail form at fixed $r>r_+$}
Along any worldline of fixed $r$ outside the KBH, each specific mode
$l,m$ decays at late time with the tail
\begin{equation} \label{Summary1}
\Psi^{slm}(t\gg |r_*|)\propto t^{-(l+l_0+3+q)}\quad
\text{for each $l\geq l_0=\max(|s|,|m|)$},
\end{equation}
to leading order in $|r_*|/t$ [see Eq.\ (\ref{tailKerr})].
Recall that $q=0$, except for $s=0$ with odd $l-l_0$ in which case $q=1$.

The most dominant modes of the {\em overall} field $\Psi^s$ are those
with $l=|s|$ and $-|s|\leq m\leq |s|$. From Eq.\ (\ref{global0}) we find,
to leading order in $|r_*|/t$,
\begin{equation} \label{Summary2}
\Psi^{s}(t\gg |r_*|)=\sum_{m=-|s|}^{|s|}
a_0^{l=|s|}\,Y^{s,l=|s|,m}(\theta,\varphi)\,\phi_r^{l=|s|}(r)\,
t^{-(2|s|+3)}  \quad  \text{(overall behavior)}.
\end{equation}
Here, the function $\phi_r^l(r)$ is the physically regular static
solution, whose exact analytic form is given in Eqs.\ (\ref{basis1}),
(\ref{phi0}), and (\ref{P}), corresponding, respectively, to the
case $s\neq 0$ with $m\neq 0$, the case $s\neq 0$ with $m=0$, and the
case $s=0$.
The constant coefficient $a_0$ (which is also $m$ dependent) is related
in Eq.\ (\ref{coefficient}) to the amplitude of the leading-order tail
at null infinity, which, in turn, is expressed as a
functional of the initial data function---see Eq.\ (\ref{I0}) and the
Appendix.
Note that Eq.\ (\ref{Summary2}) constitutes an exact analytic expression
(accurate to leading order in $|r_*|/t$)
for the late time behavior of the fields $\Psi^s$, valid {\em anywhere}
at fixed $r>r_+$.

The power law indices predicted in Eqs.\ (\ref{Summary1}) and
(\ref{Summary2})
agree with those obtained by Hod \cite{HodII} at fixed $r\gg M$ (the
result by Hod refers only to this asymptotic domain).
The result in Eq.\ (\ref{Summary2}) has support from numerical
simulations (in 2+1 dimensions) by Krivan {\em et al.}---
see Ref.\ \cite{KrivanI} for $s=0$ and Ref.\ \cite{KrivanII}
for $s=-2$.
Also, the validity of our prediction, Eq.\ (\ref{Summary1}),
has recently demonstrated numerically by Krivan \cite{Private}.

\subsubsection* {Tail form along the EH}
It is most natural to express the results at the EH in terms of
the ``physical'' fields $\hat\Psi^s\equiv \Delta^s\Psi^s$,
which are related through an EH-regular transformation to the components
of the Maxwell and Weyl tensors (see the discussion in Sec.\
\ref{secLTE}).
By virtue of Eqs.\ (\ref{EH2}), (\ref{EH3}), and (\ref{tildek2}), we
find each specific mode $l,m$ to decay along the EH with the tail
\begin{equation} \label{Summary3}
\hat\Psi^{slm}(v\gg |r_*|)\propto
\left\{\begin{array}{ll}
    v^{-(l+l_0+4+q)}& \text{for $s>0$, $m=0$},\\
    e^{im\Omega_+v}\,v^{-(l+l_0+3+q)}& \text{in all other cases},
    \end{array}
\right.
\end{equation}
(to leading order in $|r_*|/v$), with $\Omega_+$ defined in Eq.\
(\ref{Omega}).
Note that for the $s>0$ fields the axially symmetric ($m=0$) mode
decays faster than other modes. Consequently, the late time
behavior of the {\em overall} field $\hat\Psi^{s>0}$ is dominated by
the non-spherically symmetric, $m\neq 0$, modes.
These modes oscillate along the null generators of the EH with
(advanced time) frequencies $m\Omega_+$.
We find, to leading order in $|r_*|/v$,
\begin{equation} \label{Summary4}
\hat\Psi^{s}(v\gg |r_*|)=
\left\{\begin{array}{ll}
    \sum_{|m|=1}^{|s|}a_0^{l=|s|}\,Y^{s,l=|s|,m}\,(\theta,\tilde\varphi_+)
    e^{im\Omega_+v}\,v^{-(2|s|+3)}, & \text{overall $s>0$ field},\\
    \sum_{|m|=0}^{|s|}a_0^{l=|s|}\,Y^{s,l=|s|,m}\,(\theta,\tilde\varphi_+)
    e^{im\Omega_+v}\,v^{-(2|s|+3)}, & \text{overall $s\leq 0$ field},
    \end{array}
\right.
\end{equation}
where the regularized azimuthal coordinate $\tilde\varphi_+$ is the one
defined in Eq.\ (\ref{tildephi}).
Thus, the late time behavior of the field $\hat\Psi^{s>0}$ along
the EH is {\em characteristically oscillatory}.
On the other hand, the behavior of the scalar field ($s=0$)
is characteristically {\em non-oscillatory},
whereas the field $\hat\Psi^{s<0}$ involves both oscillatory
and non-oscillatory modes.

As recently discussed by Ori \cite{Ori99}, the characteristics of
the late time decay along the EH --- both the value of the power index
and the oscillatory nature of the waves --- have important implications
to the structure of the infinite blue-shift singularity at the inner
horizon of the KBH.
This singularity is related with the behavior of the ingoing component
$\Psi_0=\Psi^{s=2}$ of the Weyl perturbation. As it turns out
\cite{Ori99}, this singularity is generically oscillatory.

\subsubsection* {Tail form at null infinity}
For each specific mode $l,m$, the analysis in Sec.\ \ref{NullInfinity}
predicts at null infinity a late time tail of the form
\begin{equation} \label{Summary5}
\Psi^{slm}(u\gg M)\propto Y^{slm}(\theta,\varphi)\,v^{-2s-1}\,
u^{-(l-s+2)} \times [\text{l.d.}]
\end{equation}
to leading order in $u_0/u$ and in $M/u_0$. (We assume here that
the initial pulse is emitted at large distance, so that $-u_0\gg M$.)
In this expression, ``$[\text{l.d.}]$'' represents a possible
logarithmic dependence of the form $\ln^L[u/(r_+-r_-)]$, where
the power $L$ is some positive integer.
Such a logarithmic factor does not occur for the dominant modes $l_0$
of each $m$, and also for all modes of a scalar field ($s=0$).
Our analysis indicates that logarithmic factors do occur for the
less dominant modes ($l>l_0$) of $s\neq 0$ field; however, this point
was not studied by us in full detail.
In any case, for each given $s$ and $m$, the dominant late time decay
at null infinity is described by Eq.\ (\ref{Psil0}), in which no
logarithmic factors occur.

The {\em overall} field $\Psi^s$ is dominated at null infinity by the
modes with $l=|s|$ and $-|s|\leq m\leq |s|$:
\begin{equation} \label{Summary6}
\Psi^{s}(u\gg M)=\sum_{m=-|s|}^{|s|}
\alpha_0^{l=|s|}\,I_0^{l=|s|}\,Y^{s,l=|s|,m}(\theta,\varphi)\,
v^{-2s-1}\,u^{-(|s|-s+2)} \quad\text{(overall behavior)}
\end{equation}
to leading order in $u_0/u$ and in $M/u_0$.
Here, $I_0^{l=|s|}$ is a functional whose construction from the
initial data is described by Eqs.\ (\ref{I0}) and (\ref{g0}).

The power-law indices given in Eqs.\ (\ref{Summary5}) and (\ref{Summary6})
are in agreement with Hod's results \cite{HodI,HodII}, though Hod
indicates no logarithmic dependence for any on the modes.

\subsubsection*{Non-generic initial data}

We now briefly discuss the case where the initial pulse
is of a non-generic mode composition, such that (for
given $s$ and $m$) it does not contain the mode $l=l_0$.
For example, what can we say about a case in which, for a
$s=\pm 2$ field, the angular dependence of the initial pulse
is that of a pure mode $m=0$, $l=4$?

The calculation scheme presented in Sec.\ \ref{Global}, based on
the LTE, allows one to obtain the power index of the tail at
fixed $r$ regardless of the initial setup, provided only that the
power index at null infinity is known.
Suppose that, for a specific initial setup and for a certain $m$,
the most dominant mode, $l_0=\max(|s|,|m|)$, falls off
at null infinity with a tail of the form $\Psi^{sm}\propto u^{-w}$.
Then, from Eq.\ (\ref{sum3}) (whose derivation does not involve any
reference to the details of the initial data) we must have
$w=k_0-l_0-s-1$, where $k_0$ is the power index of the tail at
fixed $r$.
Therefore, for {\em any} initial mode composition, there exists a
simple relation between the power law indices at null infinity and
at fixed $r$. Symbolically, we may write (for each $m$)
\begin{equation} \label{Summary7}
\Psi^{sm}_{\text{scri+}}\propto u^{-w}\quad \Rightarrow \quad
\Psi^{sm}_{r={\rm const}>r_+}\propto t^{-(w+l_0+s+1)},
\end{equation}
which is valid for any initial mode composition.

The main challenge, then, is to obtain $w$, the power index
at null infinity.
Our iterative scheme, presented in Sec.\ \ref{NullInfinity},
provides a formal way for accomplishing this task;
however, in the case of a non-generic initial mode composition this
technique becomes less practical, for the following reason:
Consider, for example, an initial pulse of scalar radiation,
composed of only the mode $m=0$, $l=4$. Then, the function $\psi_{n=1}$
(namely, the first-order iteration term; we use here the notation
of Sec.\ \ref{NullInfinity}) would contain only the three modes $l=2,4,6$.
The most dominant mode of the overall field, that is $l=0$, would be
excited only at $n=2$ (and, as we suspect, will gain its typical
power-law form only at $n=3$).
Thus, in this example, it would require one to go through at least three
successive iteration stages in order to recover the tail form of the
dominant mode. (Recall that in this paper we only discussed the first
iteration; the second iteration already becomes very complicated
for analytic treatment.)
In general, as larger is the difference between the initial mode
and $l_0$, as greater becomes the number of iterations required
to extract the exact tail form of the dominant mode.
It is only for the generic case discussed in this paper that
a single iteration suffices for this goal.

The case of non-generic initial data (specifically, the case
of any pure initial mode) was studied by Hod in Refs.
\cite{HodI} and \cite{HodII}.
However, recent numerical experiments by Krivan \cite{Rapid}
show disagreement with Hod's results in this case.
Further work is needed, both analytic and numerical, to
clarify this point.

\subsubsection*{Final remarks}

We recall that in this paper we have considered only non-extremal,
$|a|<M$, Kerr BHs. Clearly, the extremal case needs to be analyzed
separately [note, for example, that Eqs.\ (\ref{r*})--(\ref{z})
cease to be valid in the case $|a|=M$].
A basic property of the effective potential in both the SBH and the
non-extremal KBH spacetimes --- its exponential decay towards the EH
(with respect to $r_*$) --- is no longer valid in the extremal Kerr case:
rather than $V(r)\propto e^{2\kappa_+ r_*}$ for $|a|<M$ near the EH,
one finds $V(r)\propto r_*^{-2}$ for $|a|=M$ [where in both cases the
tortoise variable $r_*$ is defined through the differential relation
$dr_*/dr=(a^2+r^2)/\Delta$].
Consequently, basic parts of the analysis presented in this paper
may fail to apply in the extremal case. In particular, the crucial
assumption made in Sec.\ \ref{NullInfinity}, that the late time tail at
null infinity is exclusively dominated by waves scattered at
very large distances, need not necessarily hold in this case: Here,
strong contribution to the tail may occur also from back-scattering at small
distances! To clarify the situation in the extremal case, a separate
detailed analysis is thus required.

Finally, we should comment on the limited significance
of ``multipole modes'' in the Kerr spacetime:
The spin-weighted spherical harmonic functions $Y^{slm}$
are not related here to an underlying symmetry group,
as they are in spherically symmetric backgrounds.
Consequently, the ``multipoles'' associated with these functions
have no invariant meaning, but are rather related to (and defined
through) the specific choice of the coordinates.
Yet, the functions $Y^{slm}(\theta,\varphi)$ with $\theta,\varphi$
being the Boyer-Lindquist coordinates are signified as the natural
basis for our purpose, because, at the late time limit, the field
equation becomes separable in terms of these functions.
(This separability is manifested in the $k=0$ term of the LTE,
which exhibits no coupling between the various modes $l,m$.)
Note also that the late time behavior of the overall field $\Psi^s$
is in all cases governed by a pure mode $l$.

\section*{Acknowledgement}

I wish to thank Professor Amos Ori for his guidance throughout this
research and for many helpful discussions.

\appendix

\section{Calculation of \lowercase{$\psi_{n=1}^l$} at null infinity}
\label{Appendix}

In this appendix we calculate the ``1st-order'' iteration term
$\psi_{n=1}^l$ at null infinity, at large retarded time
(that is, for $v\to \infty$ with fixed $u\gg M$).
In the case $s\geq 0$ we shall give full details of the calculation.
For brevity, the case $s<0$ (which turns out slightly more complicated
to analyze for technical reasons---see below) will be discussed in
less detail.

The starting point for the calculation would be Eq.\ (\ref{psi1Integral}),
in which the source function $S_{n=1}^l$ is given by Eq.\ (\ref{S}) as a
function of various modes of $\psi_{n=0}$.
The radial functions $\delta V(r)$, $\delta R(r)$, and $K(r)$
involved in the expression for $S_{n=1}^l$
are implicit functions of $r_*=(v-u)/2$ (because $r$ is an
implicit function of $r_*$). We can expand each of these radial functions
in powers of $1/r_*$. By virtue of Eqs.\ (\ref{AsyPotentials}) and
(\ref{DVDR}) we then obtain the leading order forms
\begin{mathletters}\label{A1}
\begin{equation} \label{A1a}
\delta V(r)=
\frac{M-imsa+l(l+1)M\left[2\ln\left(\frac{r_*}{r_+-r_-}\right)-1\right]}
{2r_*^3} + O\left[\frac{(\ln r_*)^2}{r_*^{4}}\right]
\end{equation}
and
\begin{equation} \label{A1b}
\delta R(r)=
\frac{s\left[iac_0^l-3M+2M\ln\left(\frac{r_*}{r_+-r_-}\right)\right]}
{2r_*^2} + O\left[\frac{(\ln r_*)^2}{r_*^{3}}\right],
\end{equation}
\end{mathletters}
with the asymptotic form of $K(r)$ given in Eq.\ (\ref{AsyK}).

The various terms in the source $S_{n=1}^l$ contribute additively to
$\psi_{n=1}^l$ [via Eq.\ (\ref{psi1Integral})].
The analysis below implies that the dominant contribution
to $\psi_{n=1}^l$ at null infinity at large $u$
comes only from the leading-order form (in $1/r_*$) of each of these
source terms: Roughly speaking, each additional $1/r_*$ factor in the
source leads to an additional factor of $1/u$ in the contribution
to $\psi_{n=1}^l$ at null infinity.
Hence, to calculate $\psi_{n=1}^l$ to leading order in $1/u$, one may
replace the actual functions $\delta V$, $\delta R$, and $K(r)$,
with their above asymptotic forms.

Let us now consider the contribution to $\psi_{n=1}^l$ due to a source
term of the form
\begin{equation} \label{A2}
S_{n=1}^l(u',v';l'Pd)\equiv
(2r_*)^{-P}\left[\alpha +\beta\ln (2\tilde r_*)\right]
\frac{d^d(\psi_0^{l'})}{dt^d},
\end{equation}
where a tilde symbol over a quantity shall represent the ratio of that
quantity and $(r_+-r_-)$ [so that $\tilde r_*\equiv r_*/(r_+-r_-)$].
This is the general form (to leading order in $1/r*$) of all contributing
terms in $S_{n=1}^l$, with the integer numbers $P$, $l'$, and $d$ admitting
the possible values $P=2,3$, $l'=l,l\pm 1,\pm 2$ and $d=0,1,2$, and where
$\alpha$ and $\beta$ are constant coefficients.

Let us denote the contribution of $S_{n=1}^l(u',v';l'Pd)$ to
$\psi_{n=1}^l$ at null infinity by $\psi_{n=1}^{ll'Pd}$. Then, from
Eqs.\ (\ref{psi1Integral}) and (\ref{psi0}) we find
\begin{eqnarray} \label{A3}
\psi_{n=1}^{ll'Pd}&=&\frac{v^{-s}}{(l-s)!}\sum_{j=0}^{l'-s}
A_j^{l's} \int_{u_0}^{0}du'[g_0^{l'}(u')]^{(j+d)}\int_{u}^{\infty}dv'
\frac{d^{l-s}}{du^{l-s}}\left[\frac{(v'-u)^{l+s}(u-u')^{l-s}}
{(v'-u')^{l+l'-j+P}}\right] \nonumber\\
&\times& \left[\alpha +\beta\ln (\tilde v'-\tilde u')\right].
\end{eqnarray}
Note that for $s\geq 0$, all derivatives with respect to $u$ can
be ``taken out'' of the $v'$ integration, due to the factor
$(v'-u)^{l+s}$ appearing in the integrand.
This manipulation (which is not possible for $s<0$) much simplifies
the calculation in the $s\geq 0$ case.
For brevity, we therefore continue from this point by concentrating
on the case $s\geq 0$.
Our calculations for $s<0$, whose details shall not be presented here,
yield the same qualitative results as those obtained below for $s\geq 0$,
yet they are slightly more tedious (as they involve more complicated
combinatorial expressions).
Unfortunately, we could not figure out a way for treating the $s<0$
case in a simple manner as the $s\geq 0$ case, though we think this
should be possible.

After ``taking the $u$ derivatives out'' of the integration over $v'$,
we can now easily extract the $u$ dependence of $\psi_{n=1}^{ll'Pd}$
by transforming in Eq.\ (\ref{A3}) to the new integration variable,
$x(v')=(u-u')/(v'-u')$.
This yields
\begin{eqnarray} \label{A4}
\psi_{n=1}^{ll'Pd}&=&\frac{v^{-s}}{(l-s)!}\sum_{j=0}^{l'-s}
A_j^{l's} \int_{u_0}^{0}du'\,[g_0^{l'}(u')]^{(j+d)}
\times \nonumber\\
&&\frac{d^{l-s}}{du^{l-s}}\left\{
(u-u')^{l-l'-P+1+j}\left[
\bar\alpha_j\left(\alpha +\beta\ln (\tilde u-\tilde u')\right)\,
-\bar\beta_j\beta\right] \right\},
\end{eqnarray}
where $\bar\alpha_j$ and $\bar\beta_j$ are constant coefficients
given by
\begin{eqnarray} \label{A5}
\bar\alpha_j&\equiv& \int_0^1 dx\, (1-x)^{l+s} x^{l'-s-j+P-2}=
\frac{(l'-s+P-2-j)!(l+s)!}{(l+l'+P-1-j)!},
\nonumber\\
\bar\beta_j &\equiv& \int_0^1 dx\, (1-x)^{l+s} x^{l'-s-j+P-2}\ln x.
\end{eqnarray}
[The coefficient $\bar\alpha_j$ is just the standard beta function,
$B(l'-s+P-1-j, l+s+1)$.]

Next, in Eq.\ (\ref{A4}) we integrate by parts $j+d$ successive times
with respect to $u'$. All resulting surface terms vanish due to
the compactness of $g_0^{l'}(u')$, and one is left with
\begin{eqnarray} \label{A6}
\psi_{n=1}^{ll'Pd}&=&\frac{v^{-s}}{(l-s)!}\sum_{j=0}^{l'-s}
A_j^{l's} \int_{u_0}^{0}du'\,g_0^{l'}(u')
\times \nonumber\\
&&\frac{d^{l-s+j+d}}{du^{l-s+j+d}}\left\{
(u-u')^{l-l'-P+1+j}\left[
\bar\alpha_j\left(\alpha +\beta\ln (\tilde u-\tilde u')\right)\,
-\bar\beta_j\beta\right] \right\}.
\end{eqnarray}
Here, we have used the fact that the $u'$ derivatives operate
on functions of $(u-u')$ only, to make the replacement
$\partial_{u'}\to -\partial_u$.
Evaluated at late retarded time, $u\gg -u_0$, the last
expression takes the form (accurate to leading order in $u/u_0$)
\begin{equation} \label{A7}
\psi_{n=1}^{ll'Pd}\cong I_0^{l'}\,\frac{v^{-s}}{(l-s)!}\sum_{j=0}^{l'-s}
A_j^{l's} \frac{d^{l-s+j+d}}{du^{l-s+j+d}}\left\{
u^{l-l'-P+1+j}\left[
\bar\alpha_j\left(\alpha +\beta\ln \tilde u\right)\,
-\bar\beta_j\beta\right] \right\},
\end{equation}
where $I_0^{l'}$ is the functional constructed from the function
$g_0^{l'}(u)$ according to Eq.\ (\ref{I0}).

Finally, performing the multiple differentiation in Eq.\ (\ref{A7}),
we find
\begin{equation} \label{A8}
\psi_{n=1}^{ll'Pd}\cong I_0^{l'}
\left[\lambda^{ll'Pd}\left(\alpha +\beta\ln \tilde u\right)\,
+\beta\eta^{ll'Pd}\right]\times v^{-s}\, u^{-(l'-s+P-1+d)},
\end{equation}
in which $\lambda^{ll'Pd}$ and $\eta^{ll'Pd}$ are constant coefficients.
Here, the term proportional to $\lambda^{ll'Pd}$ contains only
contributions which arise from all $l-s+j+d$ derivatives in Eq.\ (\ref{A7})
acting on the power $u^{l-l'+j-P+1}$, with non acting on $\ln\tilde u$.
The order of differentiation, $l-s+j+d$, is in all relevant
cases greater than the power index $l-l'+j-P+1$. Hence, the only
contributions to the coefficient $\lambda^{ll'Pd}$ arise when the power
$l-l'+j-P+1$ is {\em negative},
i.e.\ from the terms with $j\leq l'-l+P-2$. Since the index $j$ takes no
negative values, we find that there would
be no contribution to $\lambda^{ll'Pd}$ unless $P\geq l-l'+2$. Namely,
\begin{equation} \label{A9}
\lambda^{P<l-l'+2}=0.
\end{equation}
For $P\geq l-l'+2$, the coefficient $\lambda^{ll'Pd}$ is given by
\begin{equation} \label{A10}
\lambda^{P\geq l-l'+2}=\frac{(l+s)!}{(l-s)!}\sum_{j=0}^{\bar j}
A_j^{l's}(-1)^{l'-s+j+d}\frac{(l'-s+P-2-j)!(l'-s+P+d-2)!}{(l+l'+P-1-j)!
(l'-l+P-2-j)!},
\end{equation}
where $\bar j=\min(l'-s, l'-l+P-2)$ (in accordance with the above
discussion).
An expression for the coefficient $\eta^{ll'Pd}$ can also be written
down explicitly, using Eq.\ (\ref{A7}). It can be verified that this
coefficient is non-vanishing for all relevant values of $l,l',P,d$.

Equation (\ref{A8}) suggests that, potentially, one may expect
logarithmic dependence to occur at the leading order tail at null infinity.
In the following we show, however, that due to vanishing of the coefficient
$\lambda^{ll'Pd}$ in certain cases, this logarithmic dependence is avoided
as far as the most dominant mode of the overall field is concerned:
this dominant mode shall appear to die off with a pure power-law tail.

Note also that Eq.\ (\ref{A8}) confirms our above assertion, that to leading
order in $1/u$, the contribution from each given term in $S_{k=1}^{l'}$
(with given $l'$ and $d$) comes merely from the leading order in
$1/r'_*$: For higher-order terms in the $1/r'_*$ expansion of the source
there correspond larger values of $P$, leading to a faster decay in Eq.\
(\ref{A8}).\footnote{
Higher-order terms in the $1/r'_*$ expansion of the source
would exhibit higher $\ln \tilde r'_*$ powers, leading to higher
logarithmic powers in Eq.\ (\ref{A8}); however, the $u^{-(l'-s+P-1+d)}$
power law would remain the same for the higher order terms as well,
with $P$ denoting the power of $1/r'_*$ for each term.}

Using Eq.\ (\ref{A8}) we can now analyze the contribution to
$\psi_{n=1}^l$ at null infinity at late retarded time, belonging
to each of the various source modes $l'$:

\paragraph{Contribution of the mode $l'=l-2$:}
For this source mode we have $d=2$, and, at the leading order in $1/r'_*$,
$P=2$ and $\beta=0$ [see Eqs.\ (\ref{S}) and (\ref{AsyK})].
Since for this order $P<l-l'+2=4$, the corresponding $\lambda$ coefficient
vanishes in Eq.\ (\ref{A8}). Thus, we find no contribution
at all to $\Psi_{n=1}^l$ from the order $O\left[(r'_*)^{-2}\right]$ of
this source mode.
Turning next to the following order, with $P=3$, we find again that
$\lambda$ vanishes (as $P<4$). However, at this order the logarithmic
coefficient $\beta$ of the source does not vanish
[we have $\beta=8Ma^2C^l_{--}$---see Eqs.\ (\ref{S}) and (\ref{AsyK})],
and from Eq.\ (\ref{A8}) we find the nonvanishing, {\em nonlogarithmic}
contribution
\begin{equation} \label{A11}
\psi_{n=1}^{l-2\to l}\cong 8Ma^2C^l_{--}\,
\eta^{l'=l-2}_{P=3,d=2}\, I_0^{l-2}\, v^{-s}\,u^{-(l-s+2)},
\end{equation}
where we adopt the notation $\psi_{n=1}^{l'\to l}$ to represent the
late time contribution to $\psi_{n=1}^l$ at null infinity due to the
mode $l'$. Here, the symbol `$\cong$' stands for ``leading order in
$M/u$ and in $u_0/u$''.

\paragraph{Contribution of the mode $l'=l-1$:}
For this source mode there are two contributions: one with $d=1$
[the one proportional to $c_-^l$ --- see  Eq.\ (\ref{calI})] and one
with $d=2$ (proportional to $C_{-}^l$). Both contributions have
$P=2$ and $\beta=0$ at the leading order in $1/r'_*$. Clearly, in
view of Eq.\ (\ref{A8}), the term with $d=1$ dominates the
contribution from this mode.
Now, for the leading order, $O\left[(r'_*)^{-2}\right]$, we have
$P<l-l'+2=3$; thus the corresponding $\lambda$ coefficient vanishes.
Since for this order we also have $\beta=0$, one finds no
contribution from this order to $\psi_{n=1}^l$.
The dominant contribution would come from the next order
(with $P=3$), for which both coefficients $\lambda$ and
$\beta\cdot\eta$ are nonvanishing. Therefore, this contribution
will contain a logarithmic dependence:
\begin{equation} \label{A12}
\psi_{n=1}^{l-1\to l}\cong 8iasM\,c^l_-\, I_0^{l-1}\,
\left[\lambda^{l'=l-1}_{P=3,d=1}\left(1/2+\ln\tilde u\right)\,
+\eta^{l'=l-1}_{P=3,d=1}\right]\,v^{-s}\,u^{-(l-s+2)}.
\end{equation}

\paragraph{Contribution of the mode $l'=l$:}
There are three terms in the source $S_{n=1}^l$ which are not due
to interaction with other modes [see Eq.\ (\ref{S})].
These are (i) the term proportional to $\delta V$, for which
$P=3$ and $d=0$; (ii) the term proportional to $\delta R$, for which
$P=2$ and $d=1$; and (iii) the term proportional to $C_0^l$, with
$P=2$ and $d=2$. Clearly, the $\delta V$ term and the $\delta R$ term
contribute to $\psi_{n=1}^l$ at the same order of $1/u$, whereas
the contribution from the third term in smaller by a factor of $1/u$.
We thus concentrate on the first two terms, both of which
have $P\geq l-l'+2=2$ already at the leading order in $1/r'_*$.
Hence, the dominant contribution to $\psi_{n=1}^l$ would come
from this leading order. From Eq.\ (\ref{A8}), using Eqs.\ (\ref{A1a})
and (\ref{A1b}), we obtain for the contributions of these two terms,
\begin{mathletters} \label{A13}
\begin{eqnarray} \label{A13a}
\psi_{n=1}^{l\to l}(\text{due to $\delta V$})&\cong&
-4M\, I_0^{l}\,v^{-s}\,u^{-(l-s+2)}\nonumber\\
&\times&\left\{\lambda^{l'=l}_{P=3,d=0}\left[1-imsa/M-l(l+1)+
2l(l+1)\ln\tilde u\right]+ 2l(l+1)\eta^{l'=l}_{P=3,d=0}\right\},
\nonumber\\
\end{eqnarray}
\begin{eqnarray} \label{A13b}
\psi_{n=1}^{l\to l}(\text{due to $\delta R$})&\cong&
-2Ms\, I_0^{l}\,v^{-s}\,u^{-(l-s+2)}\nonumber\\
&\times&\left[\lambda^{l'=l}_{P=2,d=1}\left(ic_0^la/M-3
+2\ln\tilde u\right)\, +2\eta^{l'=l-1}_{P=2,d=1}\right].
\end{eqnarray}
\end{mathletters}
Now, from Eq.\ (\ref{A10}) we find
\begin{eqnarray} \label{A13.5}
\lambda^{l'=l}_{P=3,d=0}&=&\frac{s(-1)^{l-s}(l-s+1)(l+s)!}{2l(l+1)(2l+1)},
\nonumber\\
\lambda^{l'=l}_{P=2,d=1}&=&\frac{(-1)^{l-s+1}(l+s)!}{2l+1}.
\end{eqnarray}
Substituting these values in Eqs.\ (\ref{A13a}) and (\ref{A13b}),
and adding up these two equations to construct the overall contribution
from the mode $l'=l$, we find that the two logarithmic terms exactly
cancel each other. (In the scalar field case, $s=0$, each of the two
logarithmic terms vanishes independently.)
The leading order over-all contribution from the mode $l'=l$ will
therefore be {\em nonlogarithmic}, exhibiting a strict power-law
tail of the form
\begin{equation} \label{A14}
\psi_{n=1}^{l\to l}\propto M\,I_0^{l}\,v^{-s}\,u^{-(l-s+2)}.
\end{equation}

\paragraph{Contribution of the mode $l'=l+1$:}
For this mode, there contribute two terms in $S_{n=1}^l$, the ones
proportional to $c_+^l$ and to $C_+^l$.
These two terms have, respectively, $d=1$ and $d=2$, and
both have (at the leading order in $1/r'_*$) $P=2$ and
$\beta=0$. However, both coefficients $\lambda^{l'=l+1}_{P=2,d=1}$
and $\lambda^{l'=l+1}_{P=2,d=2}$
turn out to vanish, resulting in the dominant contribution to
$\psi_{n=1}^l$ coming from $P=3$.
The contribution from the term proportional to $c_+^l$
(which is $\propto u^{-(l-s+4)}\ln \tilde u$) dominates the one from
the term proportional to $C_+^l$ ($\propto u^{-(l-s+5)}\ln \tilde u$),
and one finds, in summary,
\begin{equation} \label{A15}
\psi_{n=1}^{l+1\to l}\cong 8iasM\,c^l_+\, I_0^{l+1}\,
\left[\lambda^{l'=l+1}_{P=3,d=1}\left(1/2+\ln\tilde u\right)\,
+\eta^{l'=l-1}_{P=3,d=1}\right]\,v^{-s}\,u^{-(l-s+4)}.
\end{equation}

\paragraph{Contribution of the mode $l'=l+2$:}
This mode has $d=2$ and (at the leading order in $1/r'_*$)
$P=2$ and $\beta=0$. However, the coefficient
$\lambda^{l'=l+2}_{P=2,d=2}$, as well as $\lambda^{l'=l+2}_{P=3,d=2}$,
turns out to vanish, resulting in the leading order contribution
from this mode to $\psi_{n=1}^l$ coming at $P=3$ from the term in Eq.\
(\ref{A8}) proportional to $\beta\cdot\eta$.
Hence, this contribution would be {\em nonlogarithmic}:
\begin{equation} \label{A16}
\psi_{n=1}^{l+2\to l}\cong 8Ma^2C^l_{++}\,
\eta^{l'=l+2}_{P=3,d=2}\, I_0^{l+2}\, v^{-s}\,u^{-(l-s+6)}.
\end{equation}

Equations (\ref{A11}), (\ref{A12}), (\ref{A14}), (\ref{A15}), and (\ref{A16})
describe the various contributions to the tail of $\psi_{n=1}^l$ at
null infinity from all various source modes. These results are summarized
in Eq.\ (\ref{contributions}) (in Sec.\ \ref{subIIID}) using a different
notation for the amplitude coefficients.
We point out that Eq.\ (\ref{A8}), as well as all power-law
formulas derived in this appendix, is also valid in the case $s<0$,
though with different amplitude coefficients.
(As we mentioned above, we found these coefficients to be more complicated
to calculate for $s<0$; still, the same coefficients
$\lambda$ found to vanish for $s\geq 0$, also appear to vanish in the case
$s<0$, which finally leads to the same power-law contributions.)

The following are the main conclusions that can be drawn from the
analysis in this appendix:\\
(i) In general, the dominant contribution to the late time tail of a
mode $l$ of $\psi_{n=1}$ at null infinity is due to the source modes
$l$, $l-2$, and (for $s\neq 0$) $l-1$. These contributions all have the
form $\propto u^{l-s+2}$ (multiplied by a logarithmic factor in the
$l'=l-1$ case).
Contributions due to the modes $l+1$ and $l+2$ are negligible.\\
(ii) Consequently, for given $s$ and $m$, the most dominant mode of
the field $\psi_{n=1}^{sm}$ at null infinity is the lowest
radiatable one, namely the multipole $l=l_0\equiv\min(|s|,|m|)$.\\
(iii) This mode ($l_0$) admits no contributions from lower, $l<l_0$,
multipoles, and thus, to leading order in $1/u$, it is not affected
by interaction with other modes.
Equation (\ref{A14}) then implies that this mode (and thus also
the overall field $\psi_{n=1}$) admits the strict
late time power-law tail $\propto u^{-(l_0-s+2)}$, with no
logarithmic dependence.



\end{document}